%% file: paper_arxiv1.tex
\documentclass[%
 aip,
 amsmath,amssymb,
 reprint,%
]{revtex4-2}
\input{header}
\input{acronyms}\graphicspath{{./figures/}}
\begin{document}%
%
\title{Photo\hyp induced dynamics with continuous and discrete quantum baths}
\author{Zhaoxuan~Xie}%
\affiliation{Department of Physics, Arnold Sommerfeld Center for Theoretical Physics (ASC), Ludwig-Maximilians-Universit\"{a}t M\"{u}nchen, 80333 M\"{u}nchen, Germany}%
\affiliation{Munich Center for Quantum Science and Technology (MCQST), Schellingstr. 4, D-80799 M\"{u}nchen, Germany}
\author{Mattia~Moroder}%
\affiliation{Department of Physics, Arnold Sommerfeld Center for Theoretical Physics (ASC), Ludwig-Maximilians-Universit\"{a}t M\"{u}nchen, 80333 M\"{u}nchen, Germany}%
\affiliation{Munich Center for Quantum Science and Technology (MCQST), Schellingstr. 4, D-80799 M\"{u}nchen, Germany}
\email{mattia.moroder@physik.uni-muenchen.de}
\author{Ulrich~Schollw\"ock}%
\affiliation{Department of Physics, Arnold Sommerfeld Center for Theoretical Physics (ASC), Ludwig-Maximilians-Universit\"{a}t M\"{u}nchen, 80333 M\"{u}nchen, Germany}%
\affiliation{Munich Center for Quantum Science and Technology (MCQST), Schellingstr. 4, D-80799 M\"{u}nchen, Germany}
\author{Sebastian~Paeckel}%
\affiliation{Department of Physics, Arnold Sommerfeld Center for Theoretical Physics (ASC), Ludwig-Maximilians-Universit\"{a}t M\"{u}nchen, 80333 M\"{u}nchen, Germany}%
\affiliation{Munich Center for Quantum Science and Technology (MCQST), Schellingstr. 4, D-80799 M\"{u}nchen, Germany}
\email{sebastian.paeckel@physik.uni-muenchen.de}
\begin{abstract}%
The ultrafast quantum dynamics of photophysical processes in complex molecules is an extremely challenging computational problem with a broad variety of fascinating applications in quantum chemistry and biology.
%
%
Inspired by recent developments in open quantum systems, we introduce a~\acrlong{PS-HyB} method that describes a continuous environment via a set of discrete, effective bosonic degrees of freedom using a Markovian embedding.
Our method is capable of describing both, a continuous spectral density and sharp peaks embedded into it.
Thereby, we overcome the limitations of previous methods, which either capture long\hyp time memory effects using the unitary dynamics of a set of discrete vibrational modes or use memoryless Markovian environments employing a Lindblad or Redfield master equation.
We benchmark our method against two paradigmatic problems from quantum chemistry and biology.
We demonstrate that compared to unitary descriptions, a significantly smaller number of bosonic modes suffices to describe the excitonic dynamics accurately, yielding a computational speed-up of nearly an order of magnitude.
Furthermore, we take into account explicitly the effect of a $\delta$\hyp peak in the spectral density of a light\hyp harvesting complex, demonstrating the strong impact of the long\hyp time memory of the environment on the dynamics.
\end{abstract}%
\glsresetall%
\maketitle%
\section{\label{sec:introduction}Introduction}%
\begin{figure}
    \centering
    \includegraphics[width=0.5\textwidth]{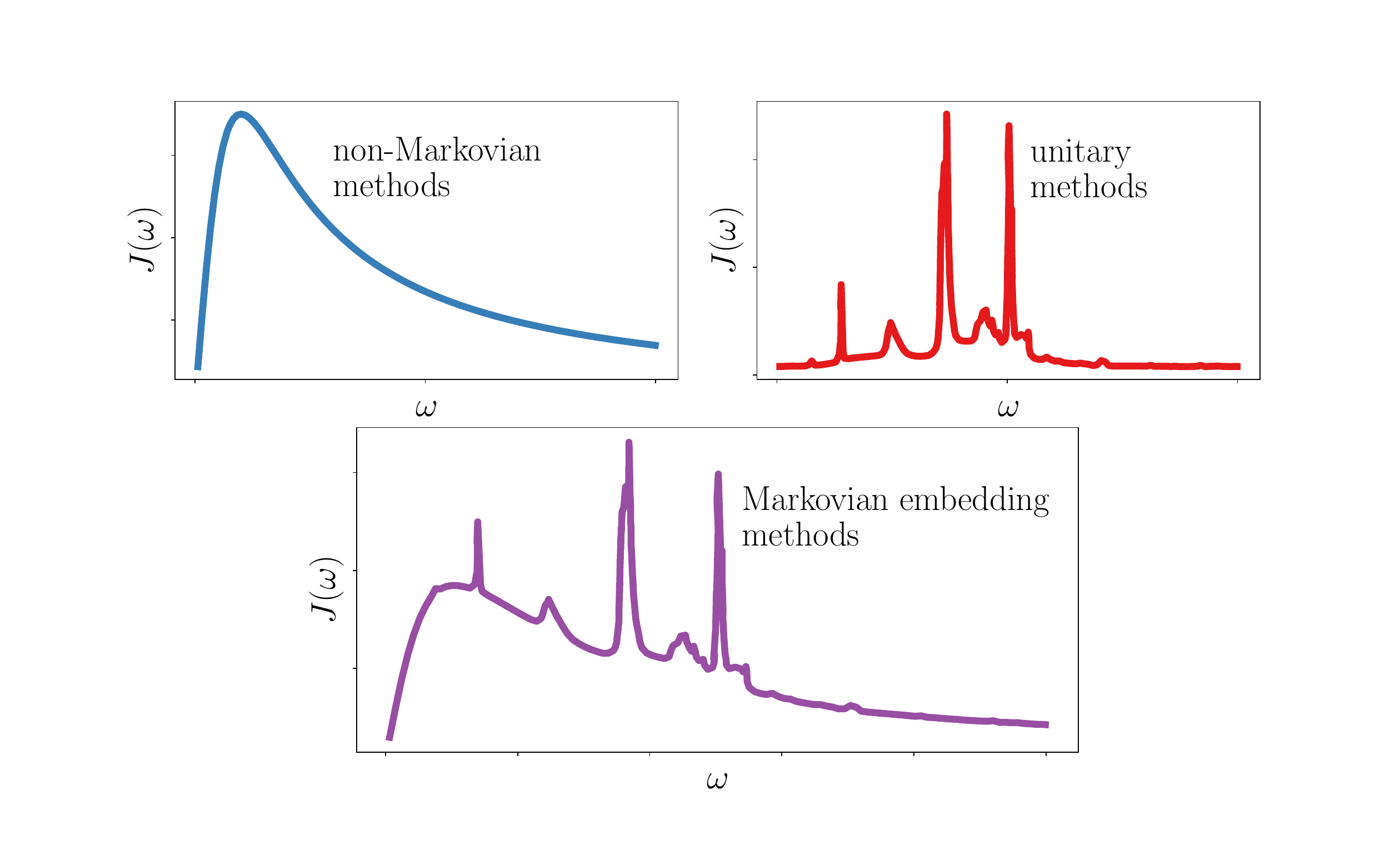}
    \caption{\label{fig:first:figure}
 Current methods for simulating the quantum dynamics of excited states coupled to vibrational~\acrlong{DOFs} target different types of environments.
 Non\hyp Markovian methods are well\hyp suited to describe continuous spectral densities (top\hyp left), while unitary methods work best for discrete spectral densities (top\hyp right), which capture long\hyp memory effects in the environment.
 We introduce a~\acrlong{PS-HyB} scheme using a Markovian\hyp embedding (bottom) to efficiently treat both types of environments.
 }
\end{figure}
The capability to simulate the ultrafast quantum dynamics of photophysical processes in complex molecules is essential to understanding phenomena such as exciton transfer and energy conversion, whose practical control is the foundation of the design of functional compounds with fascinating properties.
Their possible applications range from tailored chemical catalysts to synthetic light\hyp harvesting materials and solar cells with efficiencies beyond the Shockley\hyp Queisser limit~\cite{Xin2020,Proppe2020,10.1063/1.2356795, Congreve_science, Wang_free_triplet}.
However, the corresponding computational problems involve the dynamics of excitonic states coupled to vibrational modes, each of which is a challenging problem on its own.
Exploiting various simplifications, the computational task can be reduced to faithfully describe excitonic~\gls{DOFs} representing electronic excited states, which are (non\hyp locally) coupled to a manifold of vibrational~\gls{DOFs}.
Here, especially the vibrational system of complex molecules poses various numerical challenges and there has been a great effort to deal with the exponential scaling of the computational complexity.
In general, two conceptually different approaches were established.
(i) A discretization of the vibrational~\gls{DOFs} is combined with selection rules to choose the most relevant modes, such that the unitary dynamics of the resulting, closed quantum system can be described using efficient state representations and integration schemes~\cite{white1992density, schollwock2005density,schollwock2011density,haegeman2016unifying, paeckel2019time}.
(ii) The excitonic~\gls{DOFs} are treated exactly while the vibrational~\gls{DOFs} are traced out and encoded in an effective environment.
The complexity of the resulting, open quantum\hyp system dynamics then mainly depends on the choice of the environment~\cite{Mirjani2014-ux, nalbach2011exciton}.
The dynamics in the presence of memoryless, Markovian environments can be treated efficiently using Lindblad master equations\cite{Lindblad1976-po,Pearle_2012,dalibard1992wave, daley2014quantum, moroder2023stable}.
Structured, non\hyp Markovian environments introduce time-non\hyp local correlations between system and environment, which increase the computational cost~\cite{makri1995tensor1,makri1995tensor2,tanimura1989time,suess2014hierarchy,imamog1994stochastic, Garraway1997, PhysRevB.86.125111, 10.1063/1.5000747, PhysRevA.101.050301, brenes2020tensor, lacerda2023quantum, lacerda2023entropy} but are crucial to simulate the inherently mesoscopic vibrational systems of molecules, exhibiting long\hyp time memory effects.
To efficiently simulate the quantum dynamics of both, coupled electronic and vibrational, as well as environmental \gls{DOFs}, the decomposition of the many\hyp body wave function into local tensors has proven to be an extremely fruitful approach\cite{MEYER199073, xie2019time, Shi2018,strathearn2018efficient, Zheng2016-ju, Broch2018-vl, Reddy2018-ft, Schulze2016, Jiang2020}.
The unitary evolution of excitonic states coupled to $\mathcal{O}(100)$ discrete vibrational~\gls{DOFs} is a remarkable milestone facilitating, for instance, the unbiased investigation of ultrafast photo\hyp induced dynamics of large molecules~\cite{Schroeder2019,Mardazad2021, Xie2019-ce}.
Recently, wave function\hyp based~\gls{TN} methods for open quantum\hyp systems with both Markovian and non\hyp Markovian environments have been developed~\cite{Hartmann2017-ke,PhysRevLett.128.063601,Gao2022} to describe continuous and smooth spectral densities.
However, these approaches leave a methodological gap, if the system's spectral density contains both aspects, discrete peaks immersed into a smooth background.
This situation, generated for instance by non\hyp vanishing memory effects, is inevitable in realistic materials, such as organic semiconductors, which involve both dispersive modes with continuous spectral densities and local vibrational modes with sharp peaks.
While in these cases the unitary approach would require a daunting amount of discrete vibrational modes, generic open\hyp quantum system methods become unstable in the presence of discrete peaks~\cite{HEOMrminstab}.
To fill this gap, pseudomode approaches have been introduced to approximate the spectral density by introducing damped auxiliary modes~\cite{Garg1985,imamog1994stochastic,Garraway1997,Dalton2001,Tamascelli2018} and solving the equations of motion for the density operator~\cite{Plenio2019,Mascherpa2020}.
Here, we propose a~\gls{TN}\hyp based~\gls{PS-HyB} scheme computing quantum trajectories in terms of~\gls{MPS} of an enlarged system with Markovian embedding, where both continuous and discrete contributions to the spectral density can be handled (see ~\cref{fig:first:figure}).
Simulating two paradigmatic examples, i.e., singlet fission in a three\hyp state model with Debye spectral density, as well as finite\hyp temperature exciton\hyp dynamics in a~\gls{FMO} complex with an additional sharp peak in the spectral density, we demonstrate that~\gls{PS-HyB}\hyp\gls{MPS} can efficiently capture the regime of long\hyp time memory effects at significantly smaller computational cost than unitary methods, while it circumvents instabilities from a direct evolution of the density operator.
This is achieved by exploiting recent advances for~\gls{TN} representations of systems with large local Hilbert spaces, rendering a pure\hyp state unraveling practically feasible, also in the presence of strong exciton\hyp phonon couplings~\cite{10.21468/SciPostPhys.10.3.058, STOLPP2021108106,Mardazad2021, Xu2023-pl,moroder2024phonon}.
\section{\label{sec:methodology}Methodology}
We aim to describe the dynamics of $L$ excitonic states, each interacting with a set of vibrational~\gls{DOFs}, acting as independent environments.
For this system\hyp environment partitioning, the spectral density $J^I(\omega)$ ($I = 1, 2, \dots L)$ is the key quantity encoding the relevant information about the coupling between excitonic and the vibrational~\gls{DOFs}.
Our goal is to capture the general situation of a continuous background and discrete, sharp peaks describing long\hyp time memory effects, as pictured in~\cref{fig:first:figure}.
For that purpose, we adopt a method called mesoscopic leads or extended reservoirs\cite{imamog1994stochastic, Garraway1997, PhysRevB.86.125111, 10.1063/1.5000747, PhysRevA.101.050301, brenes2020tensor, lacerda2023quantum}.
In this framework, the general idea is to approximate each bath spectral density $J^I(\omega) = \pi/2 \sum_k |g^I_k|^2 \delta(\omega-\omega_k)$ by fitting it with $N^I$ Lorentzians:
\begin{equation}
 J^I(\omega) = \sum_{m=1}^{N^I} \frac{ \gamma^I_m \abs{g^I_m/2 }^2 } {(\omega-\omega^I_m)^2 + (\gamma^I_m/2)^2} \;,
    \label{eq:spectral:density:fitting}
\end{equation}
where the couplings between excitonic and vibrational~\gls{DOFs} are given by $g^I_m = \sqrt{ 2 J^I(\omega^I_m) \gamma^I_m/\pi}$, \textcolor{red}.
Each Lorentzian contributing to the spectral density can be represented by a harmonic oscillator with frequency $\omega^I_m$ coupled to a memoryless environment, where the spectral broadening is given by the potentially non\hyp uniform discretization $\gamma^I_m$.
This way, we can represent a continuous spectral density with sharp features by a set of $\sum_I N^I$ bosonic modes coupled to Markovian environments.
\subsection{\label{sec:methodology:ps-hyb}\Acrlongpl{PS-HyB}}
\begin{figure}
    \centering
    \includegraphics[clip,trim=1.5em 1.5em 4.9em 4.4em,width=0.48\textwidth]{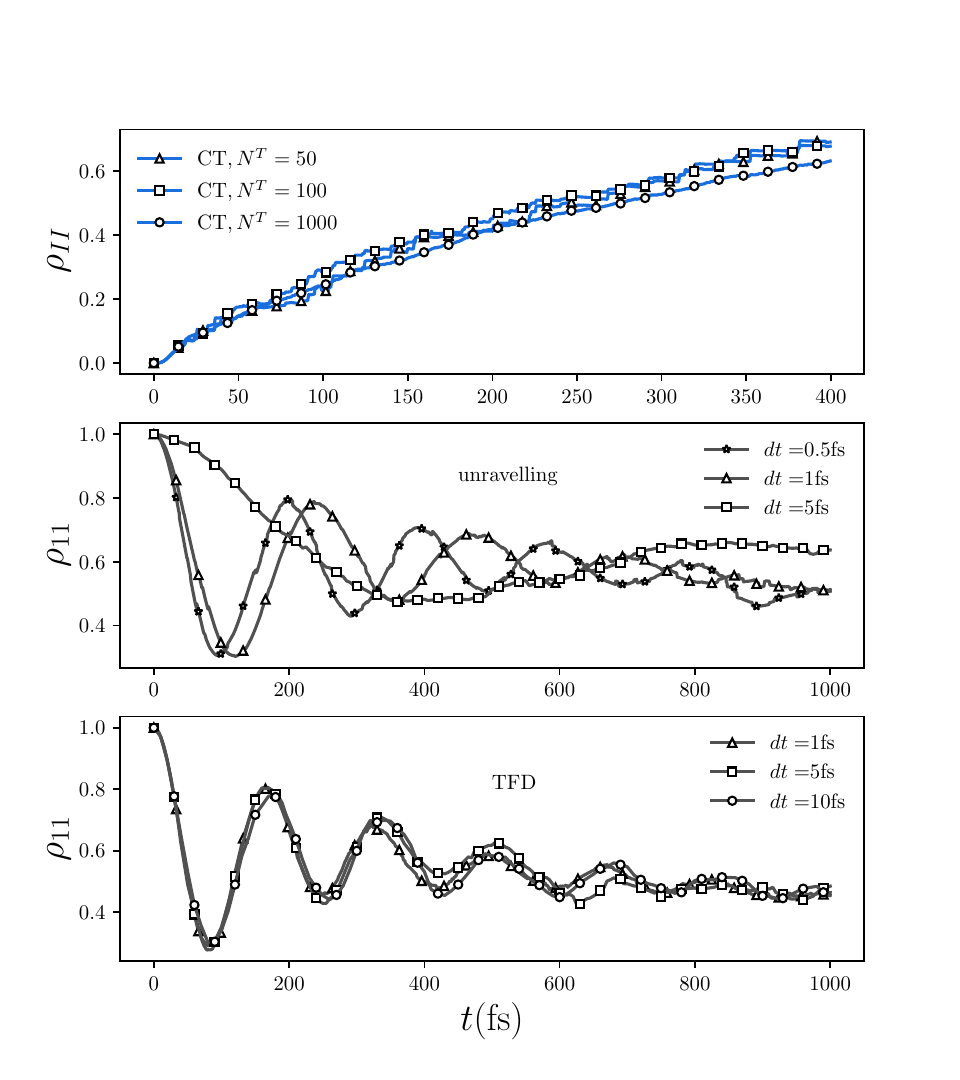}
    \caption{
    \label{fig:convergence}
    \discussive{(Top) The convergence of~\gls{PS-HyB}\hyp\gls{MPS} with the number of trajectories at the example of the~\acrshort{CT}\protect\hyp state occupation for the minimal model of singlet fission studied in~\protect\cref{sec:sf}.
    (Middle and bottom) We compare the convergence of occupation of site 1 in the FMO complex at finite temperatures using either a naive pure\hyp state unraveling of the thermal density operator (middle) or the thermofield and thermal vacuum representation described in~\protect\cref{sec:methodology:thermofield}.
    Note that when using the naive pure state unraveling, we observe a very slow convergence, requiring extremely small time steps $\delta t \leq 0.5\mathrm{fs}$.
    Instead, the thermofield and thermal vacuum representation exhibits a significantly increased stability, allowing for computationally beneficial, larger time steps $\delta t \geq 5\mathrm{fs}$.}
    }
\end{figure}

Let us now work with the explicit Hamiltonian for the combined excitonic and vibrational system
\begin{equation}
    \hat{H} = \hat{H}_\mathrm{exc} + \hat{H}_{\mathrm{vib}} + \hat{H}_{\mathrm{ex-vib}}\; .
    \label{eq:total:ham}
\end{equation}
Here, $ \hat{H}_{\mathrm{exc}}$ is an arbitrary excitonic Hamiltonian. In the following, we set $k_B, \, \hbar = 1$.
The vibrational modes are described by harmonic oscillators with frequencies $\omega^I_m$: $\hat{H}_{\mathrm{vib}} = \sum_{m,I} \omega^I_m \hat{b}^{I \dagger}_m\hat{b}^I_m$, where $ \hat{b}^{I (\dagger)}_m$ are the bosonic annihilation (creation) operators. 
The interaction between excitons and vibrational modes takes the form $ \hat{H}_{\mathrm{ex-vib}} =  \sum_{m,I} g^I_m \hat{n}^I ( \hat{b}^I_m + \hat{b}^{I \dagger}_m )/\sqrt{2}$.
Upon introducing a sufficient number of oscillators $N^I$, the dynamics of the combined system obeys the Lindblad equation~\cite{Lindblad1976-po,Pearle_2012}
\begin{equation}
    \frac{\mathrm{d} \hat \rho}{\mathrm{d}t} = -i \comm{\hat{H}}{\hat{\rho}} + \sum_I \mathcal{D}^I (\hat \rho)\; .
    \label{eq:mesoleads:lindblad}
\end{equation}
The sum on the right-hand side models the interaction of the system with a Markovian environment.
In our case, these terms describe the thermalization of each vibrational subsystem $I$ via the Davies map~\cite{DAVIES1979421,ROGA2010311}
\begin{equation}
    \mathcal{D}^I(\hat\rho) = \sum_{m=1}^{N^I} \gamma^I_m \left[ f^I_m \mathcal D(\hat{\rho},\hat L_1) + (1+f^I_m) \mathcal D(\hat{\rho},\hat L_2) \right],
    \label{eq:dissipator:mesoleads}
\end{equation}
where the thermalization rate is proportional to the spacing and $f^I_m(\beta) = 1/ ( \exp(\beta\omega^I_m)-1 )$ is the Bose-Einstein distribution function at inverse temperature $\beta$.
Here, we introduced the dissipator
\begin{align}
    \mathcal D(\hat \rho, \hat L_l) = \hat L^\nodagger_l \hat \rho \hat L^\dagger_l - \frac{1}{2}\left\{\hat L^\dagger_l \hat L^\nodagger_l, \hat \rho \right\} \; ;
\end{align}
specifying the couplings to the environment via the jump operators $\hat L_1 = \hat b^{I \dagger}_m$ and $\hat L_2 = \hat b^{I}_m$.
For a large number of bath modes $N^I$, the dissipative evolution of the excitonic subsystem~\cref{eq:mesoleads:lindblad} becomes equivalent to the unitary evolution with discretized bath spectral densities~\cite{devega2015discretize}, described for instance via a chain mapping \cite{lacerda2023quantum}.
We will show the mesoscopic leads approach to be numerically much more efficient when solving the dynamics using~\gls{TN} methods.
This is based on the observation that one obtains the unitary case from taking the zero\hyp broadening limit of~\cref{eq:mesoleads:lindblad}.
In fact, when $J(\omega)$ is continuous and smooth (~\cref{fig:first:figure}(top left)) a large number of discrete modes would be required to accurately fit $J(\omega)$.
Moreover, unitary dynamics suffer from severe finite\hyp size errors caused by the scattering of excitations at the systems' boundaries, which limits the reachable simulation time~\cite{lacerda2023quantum}.
Nonetheless, compared to unitary time evolution, Lindblad dynamics pose challenges to numerical treatments not only because of the dimension but also because the truncation of a mixed state requires particular care~\cite{cui2015variational, werner2016positive}.
Thus, it is convenient to resort to so\hyp called pure\hyp state unravelings, which consist of reconstructing the Lindblad dynamics from an ensemble of stochastic pure\hyp state time evolutions.
Different pure\hyp state unraveling schemes, such as \gls{QSD}~\cite{gardiner00} and the \gls{QJ}~\cite{daley2014quantum, moroder2023stable} approaches,  have been developed.
Here, we apply the \gls{QJ} scheme, which is the more widely used variant due to its good stability and convergence properties.
The starting point for the \gls{QJ} scheme is to rewrite~\cref{eq:mesoleads:lindblad} as
\begin{equation}
    \frac{\mathrm{d}\hat{\rho}(t)}{\mathrm{d}t} = -i \comm{\hat{H}_{\mathrm{eff}}}{\hat{\rho}(t)} + \sum_l \hat L^\nodagger_l \hat{\rho}(t)\hat{L}^\dagger_l \; ,
    \label{eq:lindblad:qj:heff}   
\end{equation}
where $\hat{H}_{\mathrm{eff}} = \hat{H} -i/2 \sum_l \hat L^\dagger_l \hat L^\nodagger_l$ is an effective, non\hyp Hermitian Hamiltonian.
If the initial state is pure ($\hat{\rho}(0) = \ket{\Psi(0)}\bra{\Psi(0)}$) the first term of~\cref{eq:lindblad:qj:heff} describes a simple (non\hyp unitary) Schrödinger evolution.
The action of the second term can be mapped to the stochastic application of $\hat{L}_l$ to a pure state.
\discussive{
We can achieve this by introducing a collection of random numbers defining trajectories $q(t)$ (see for instance~\onlinecite{daley2014quantum}) such that at any time $t$ the ensemble average $\mathcal{E}$ over $N^T$ pure\hyp state evolutions yields an approximation of the Lindblad-evolved density matrix $\mathcal{E}_{N^T} \left[ \ket{\Psi(t)_q}\bra{\Psi(t)_q} \right ] = \hat{\rho}_{N^T}(t)$ (which formally becomes exact for an infinite number of trajectories, i.e. $\hat{\rho}_{N^T}(t) \xrightarrow{{N^T}\to \infty} \hat{\rho}(t)$).
Usually, one avoids computing the density matrix explicitly and evaluates the expectation value of an observable of interest directly as $\langle \hat{O} \rangle_{N^T} (t) =  \mathcal{E}_{N^T} [ \bra{\Psi(t)_q} \hat{O} \ket{\Psi(t)_q} ] $.
The quick convergence of local observables in the number of trajectories is shown in ~\cref{fig:convergence}(top) as the example of a minimal model for singlet fission, which is discussed in more detail in~\cref{sec:sf}.
Using the independence of different trajectories, one can estimate the statistical error $\epsilon_{O_{N^T}}$ for the expectation value of an observable $\hat{O}$ as~\cite{daley2014quantum}
\begin{equation}
    \epsilon_{O_{N^T}}= \frac{ \sqrt{\braket{\hat{O}^2}_{N^T} - \braket{\hat{O}}_{N^T}^2} }{\sqrt{N^T}} \;.
    \label{eq:qj:error}
\end{equation}
We will adopt~\cref{eq:qj:error} as error bars for all~\gls{QJ} results.
}

\discussive{We note that~\gls{DAMPF}, an approach similar to the one discussed in this manuscript, was introduced in Ref.~\onlinecite{Plenio2019}.
\gls{DAMPF} is based on fitting the bath correlation function (which is connected to $J(\omega)$ by a Fourier transform) in the time domain and on the direct propagation of the density matrix.
While the direct propagation of the density matrix using~\gls{DAMPF} avoids the sampling error present in the~\gls{QJ} method, it can suffer from positivity issues~\cite{Eisert2014} and the evolution of matrix-density operators is in general numerically less efficient than advanced time\hyp evolution schemes for~\gls{MPS}, which we use to compute the~\gls{QJ} trajectories.
Essentially, the advantage of~\gls{PS-HyB}\hyp\gls{MPS} is based on the computationally beneficial properties of~\gls{MPS}, which are made accessible by the pure state unraveling and allow for a taylored~\gls{MPS} representation of the vibrational degrees of freedom and to invoke high\hyp performance time\hyp evolution schemes for~\gls{MPS}, which we discuss in~\cref{sec:methodology:pp-mps}.
}
\subsection{\label{sec:methodology:pp-mps}\Acrlongpl{PP-MPS}}
\discussive{
We represent the excitonic-vibronic system as a \glspl{MPS}~\cite{white1992density, schollwock2011density}.
In general, this allows to employ efficient truncation schemes for $\ket{\Psi}$ that are controlled by the bond dimension and/or the truncated weight.
However, representing the vibrational modes as an \gls{MPS} is an arduous task because one has to truncate the formally infinite bosonic Hilbert space.
Moreover, the interaction Hamiltonian breaks the $U(1)$ symmetry for the bosonic excitations.
We can address both challenges by working with \gls{PP-MPS}~\cite{10.21468/SciPostPhys.10.3.058, STOLPP2021108106}, which were successfully applied to simulate unitary photo\hyp induced dynamics in different setups~\cite{Mardazad2021, Xu2023-pl,moroder2024phonon}.
}
\discussive{
The idea of~\gls{PP-MPS} is to pair every physical bosonic degree of freedom $\ket{n_m}$ with a fictitious bath mode $\ket{n_{B;m}}$, where $n_m, n_{B;m}\in\left\{0, \ldots, d-1 \right\}$ and $d$ denotes the cutoff dimension of the bosonic Hilbert spaces.
The exchange of excitations between physical and bath degrees is constrained by imposing local gauge constraints $n_m+n_{B;m} = d-1$, which can be satisfied easily when preparing the initial state of the time evolution.
In order to conserve the local gauge constraints throughout the dynamics, bosonic excitations of the physical degrees of freedom have to be moved from or to the bath sites.
This is easily achieved by introducing balancing operators $\hat \beta^{(\dagger)}_{B;m}$ that counteract the change of excitations in the physical, bosonic system.
The local gauge constraints have various advantageous properties.
They introduce an artificial, global $U(1)$ symmetry because for each pair of physical and bath degree of freedom, the total number of excitations is conserved, which allows the employment of efficient, block\hyp diagonal representations of site tensors.
As a consequence, the representation of a local bosonic operator is transformed from one $d$\hyp dimensional block to $d$ one\hyp dimensional blocks and one can exploit efficient parallelization schemes over different quantum number sectors.
Even more importantly, it can be shown that under the introduction of the gauge constraints, the diagonal elements of the bosonic~\gls{1RDM} $\rho_{n^\noprime_m, n^\prime_m}$ are directly related to the Schmidt values when introducing a bipartition between site $m$ and $m+1$.
This allows to directly truncate at the level of the bosonic~\gls{1RDM} by using only standard~\gls{DMRG} truncation schemes.
In fact, if the probabilities $\rho_{n_m,n_m}$ to have $n_m$ excitations at vibrational mode $m$ are below a certain truncation threshold $\delta > \rho_{n_m,n_m}$, discarding these matrix elements implies an actual reduction of the local Hilbert space dimension $d \rightarrow d_\mathrm{eff}$~\cite{10.21468/SciPostPhys.10.3.058}.
In practice, this allows us to choose $d$ rather large since the required local Hilbert space dimension $d_\mathrm{eff}$ is controlled by tuning the truncation threshold $\delta$ of excitation probabilities for the physical, bosonic degrees of freedom.
The resulting truncation of the local Hilbert space dimension is similar to the effect of a local basis optimization~\cite{Zhang1998,Brockt2015}.
However, in~\gls{PP-MPS}, there is no need for an additional basis transformation on the local bosonic Hilbert spaces, which alleviates typical problems of dynamical basis adaptions and truncations, such as reintroducing previously discarded basis states~\cite{STOLPP2021108106}.
}
\discussive{
To calculate the time-evolution of a \gls{MPS}, one of the most efficient methods is a~\gls{TN} variant of the~\gls{TDVP}~\cite{haegeman2011time, haegeman2016unifying, paeckel2019time}.
It is based on the Dirac-Frenkel variational principle~\cite{Lubich2004OnVA} and yields an approximation to the solution of the Schr\"odinger equation projected to the tangent space of the variational manifold generated by the state parametrization (here the~\gls{MPS} tensor elements).
The numerical efficiency of~\gls{TDVP} in its~\gls{TN} formulation stems from the fact that the projector to the tangent space of the~\gls{MPS} manifold can be split into local projectors and thereby induces a local integration scheme that can be solved efficiently~\cite{haegeman2016unifying}.
This comes at the cost of projection errors typical for~\gls{TN} algorithms, which originates from the restriction of the local state space and can introduce uncontrolled errors.
Consequently, convergence with respect tp the~\gls{MPS} bond dimension is crucial, and a careful analysis for the systems simulated in this article can be found in~\cite{supp_mat}.
}
\discussive{
It should be noted that the so\hyp called projector\hyp splitting constitutes the major difference between classical~\gls{ML-MCTDH} implementations and the~\gls{TN} formulation of~\gls{TDVP}.
Moreover, a projector\hyp splitting for~\gls{ML-MCTDH}~\cite{Klos2017} was introduced recently avoiding the inversion of potentially ill\hyp conditioned transformation matrices~\cite{Lindoy2021a,Lindoy2021b}.
The remaining conceptual difficulty is how to deal with projection errors, which are generated by lowest\hyp rank approximations of the wavefunction.
For bosonic degrees of freedom, we extended the single\hyp site variant of~\gls{TDVP} with a local subspace expansion \cite{PhysRevB.102.094315, grundner2023cooperpaired}, which exhibits the computationally most favorable scaling to resolve this issue.
It combines single\hyp site \gls{TDVP} updates with local basis expansions, increasing the states' bond dimension and thereby reducing projection errors such that they are not the dominant source of computational errors.
The resulting expansion scheme captures the spread of correlations during the time evolution, while the dominating contractions scale only quadratically in the cutoff dimension of the local bosonic Hilbert spaces.
}
\subsection{\label{sec:methodology:thermofield}Thermofield doubling and thermal vacuum representation}
\discussive{
To describe vibrational environments at finite temperatures, we perform a thermofield doubling of the bosonic system and map the bosonic thermal state to the vacuum via a Bogoljubov transformation, as detailed in Ref.~\onlinecite{devega2015thermofield}.
This allows for an efficient representation of thermal fluctuations, if the vibrational modes are described by independent harmonic oscillators and the coupling to the excitonic system is linear.
Here, the idea is to introduce a new set of bosonic operators $\hat{a}_k$, linked to the original operators $\hat{b}_k$ by a unitary transformation, which annihilate the thermal state $\hat{\rho}^\beta_\mathrm{vib}$.
By transforming the Hamiltonian to the new basis, the thermal state can be represented as a vacuum state.
This not only avoids thermal sampling of initial states, but also most importantly, reduces the number of quantum jumps from the environment to the vibrational modes.
Thereby, the convergence of the~\gls{QJ} method is strongly improved, which is shown exemplarily in ~\cref{fig:convergence} (middle and bottom), where we also compare the convergence against a sampling of the thermal state using an additional pure\hyp state unraveling of the density operator~\footnote{For a more detailed comparison of the different finite\hyp temperature approaches see~\cite{supp_mat}}.
}
\begin{figure}
    \centering
    \subfloat[\label{fig:example:sf}]{
        \includegraphics[clip,trim=1.5em 0 4.9em 4.4em,width=0.43\textwidth]{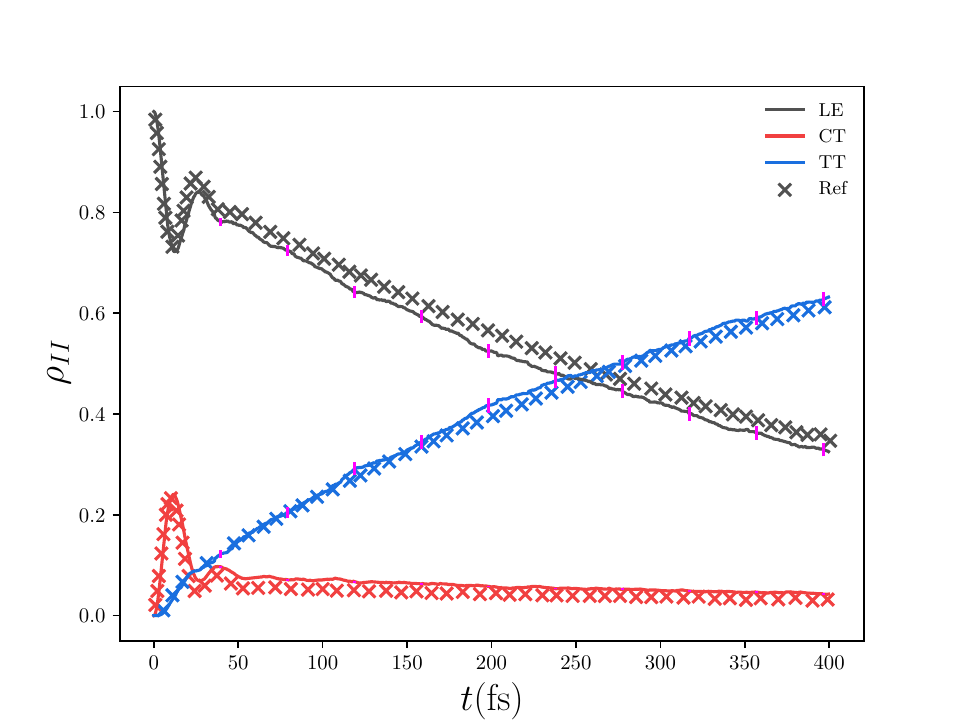}
 }

    \vspace*{-2em}
    \subfloat[\label{fig:example:fmo:300k}]{
        \includegraphics[clip,trim=1.5em 0 4.9em 4.4em,width=0.43\textwidth]{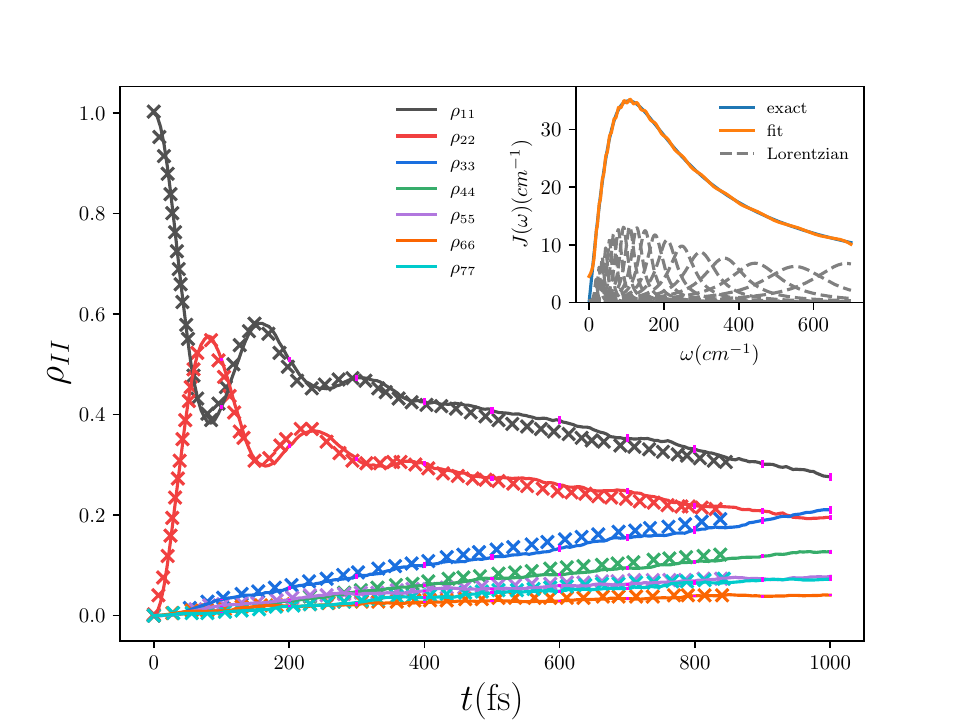}
 }

    \vspace*{-2em}
    \subfloat[\label{fig:example:fmo:77k+peak}]{
        \includegraphics[clip,trim=1.5em 0 4.9em 4.4em,width=0.43\textwidth]{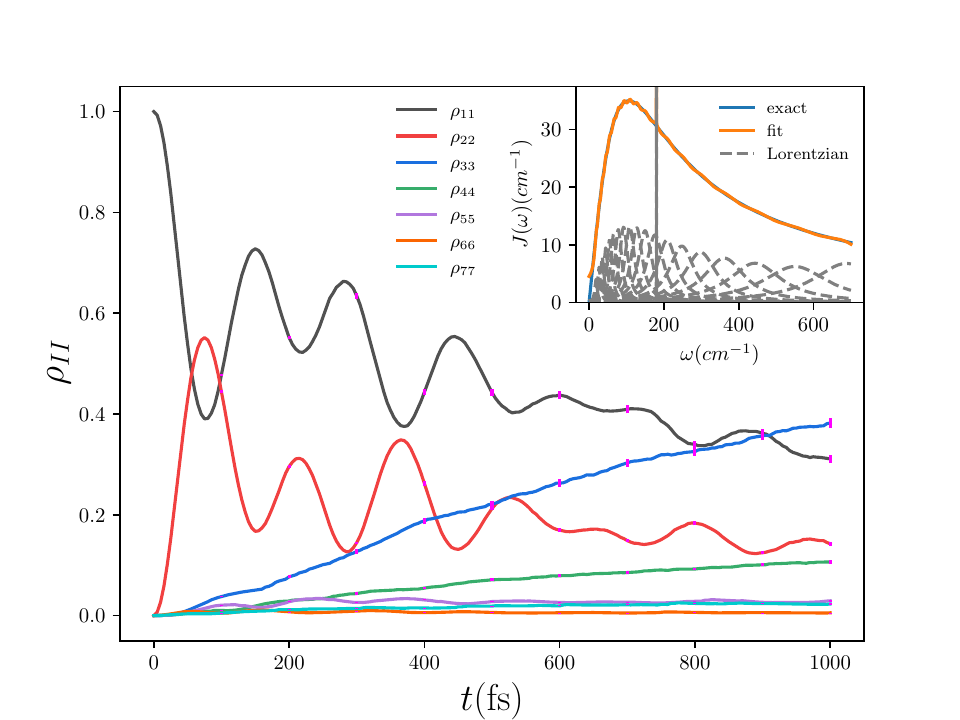}
 }
    \caption{
        \label{fig:examples}
 Dynamics of the excitonic populations $\rho_{II}=\braket{I|\hat \rho|I}$ for a three\hyp state model of~\gls{SF}~\subfigref{fig:example:sf} at $T=0\mathrm{K}$ and a photo\hyp excited~\gls{FMO} complex at $T=300\mathrm{K}$~\subfigref{fig:example:fmo:300k} and $T=77\mathrm{K}$~\subfigref{fig:example:fmo:77k+peak}, using a Debye spectral density, which in~\subfigref{fig:example:fmo:77k+peak} is extended by a vibrational mode of the~\gls{FMO} pigments.
 The solid lines are obtained using~\gls{PS-HyB}\hyp\gls{MPS} with a~\gls{QJ} pure state unraveling using $N=1000$ independent trajectories; error bars are shown only every $50 \mathrm{fs}$~\subfigref{fig:example:sf} and $100 \mathrm{fs}$~\subfigref{fig:example:fmo:300k},~\subfigref{fig:example:fmo:77k+peak}.
 Reference values are denoted by crosses, extracted from Ref.~\onlinecite{Zheng2016-ju} for~\subfigref{fig:example:sf} and Ref.~\onlinecite{nalbach2011exciton} for~\subfigref{fig:example:fmo:300k}.
 In the insets, we display the fitting of the spectral density for the example of~\gls{FMO}.
 We find excellent agreement with the reference data using a rather small number of Lorentzians $N^I=10$ for~\subfigref{fig:example:sf} and $N^I=19$ for~\subfigref{fig:example:fmo:300k} and ~\subfigref{fig:example:fmo:77k+peak}.
 }
\end{figure}
\section{\label{sec:sf}Example 1: Singlet fission}
To benchmark the capability of~\gls{PS-HyB}\hyp\gls{MPS} to reduce the necessary complexity of a faithful description of vibrational subsystems, we first studied a paradigmatic, minimal model for singlet fission at zero temperature~\cite{Zheng2016-ju,xie2019time,Xu2023-pl}.
The excitonic part of the Hamiltonian consists of $L=3$ excitonic states usually referred to as the (i)~\gls{LE}, (ii)~\gls{CT}, and (iii)~\gls{TT} states.
Each state is coupled to its own bath of harmonic oscillators characterized by a Debye spectral density
\begin{equation}
 J^D(\omega) = \frac{2 \lambda \omega \omega_c}{(\omega^2 + \omega_c^2)} \; .
    \label{eq:debye:spectral:density}
\end{equation}
We set $\lambda=807 \mathrm{cm}^{-1}$ and $\omega_c = 1452 \mathrm{cm}^{-1}$ and draw the explicit model parameter of $\hat H_\mathrm{exc}$ from~\cite{Zheng2016-ju}.
The details of the numerical implementation together with a convergence analysis can be found in~\cref{app:numerical:details}.
Following~\cref{sec:methodology}, for each excitonic state we fit the corresponding spectral density $J^D(\omega)$ and we found $N^I=10$ vibrational modes to be sufficient to faithfully reproduce $J^D(\omega)$.
At $t=0$, we prepared the excitonic system in the~\gls{LE} state and the vibrational environment in the $T=0\mathrm K$ vacuum state.
Choosing a timestep $\delta t=5 \; \mathrm{fs}$ and a cutoff dimension $d=16$ for the bosonic sites, we evaluated the time\hyp dependent excitonic populations, averaging over $1000$ independent trajectories.
In~\cref{fig:example:sf}, we show the resulting dynamics (solid lines, error bars indicated every $50\mathrm{fs}$) compared to reference data (crosses) extracted from Ref.~\onlinecite{Zheng2016-ju}.
We find an excellent agreement also at very long simulation times using only $N^I=10$ bosonic modes per excitonic state, which should be compared to $N^I=61$ modes per excitonic state used in the unitary time\hyp evolution scheme in Ref.~\onlinecite{Zheng2016-ju}.
Small deviations can be attributed to the number of sample trajectories, which is discussed in more detail in~\cite{supp_mat}.
\section{\label{sec:fmo}Example 2: Light-harvesting complex}%
Simulating complex excitonic dynamics at finite temperatures is an even more challenging problem, particularly when using wave function\hyp based methods.
As a second example, we therefore studied the photo\hyp induced dynamics of the~\gls{FMO}\hyp complex at finite temperature.
It is a pigment\hyp protein complex found in green sulfur bacteria, which is involved in the early stages of photosynthesis~\cite{FMO_Original}.
Following the literature, we used $L=7$ excitonic states for the~\gls{FMO} complex. Each state is coupled to an independent vibrational environment~\cite{nalbach2011exciton}, which we initialized at a finite temperature~\cite{devega2015thermofield} (see~\cref{app:tfd:vs:unravel}).
The excitonic Hamiltonian is given by
\begin{equation}
    \hat H_\mathrm{exc} = \sum_{I,J=1}^7 t_{IJ} \ket{I}\bra{J} \;,
\end{equation}
where we adopted the couplings $t_{IJ}$ from Ref.~\onlinecite{nalbach2011exciton}.
The standard approach to describe the vibrational system is by assuming a Debye spectral density~\cref{eq:debye:spectral:density}, which we fit with $N^I=19$ Lorentzians per vibrational environment.
We extend this approach by adding a sharp peak at a frequency $\omega_0=29\mathrm{cm}^{-1}$, which is generated from a vibrational mode of the individual pigments.
We emphasize that this modification introduces long\hyp time memory effects in the environment that can significantly alter the photo\hyp induced dynamics.
Moreover, previous studies that used a non\hyp Markovian description of the environments, only considered finite broadenings~\cite{nalbach2011exciton} of this mode, drastically reducing memory\hyp effects.
We performed simulations of the photo\hyp induced dynamics at the practically relevant temperatures $T=77$ and $300\mathrm{K}$ measuring the time\hyp dependent exciton population, starting from an initial exciton configuration $n_1=1, n_{I>1}=0$, and using $1000$ independent trajectories.
In~\cref{fig:example:fmo:300k}, we show the resulting dynamics for the case of a pure Debye spectral density, i.e., without the additional peak at $\omega_0$, at $T=300\mathrm{K}$.
The solid lines represent the time\hyp dependent excitonic state occupations and the error bars are now indicated every $100\mathrm{fs}$.
We again find excellent agreement with the reference data (crosses), reaching a very long simulation time of $1\mathrm{ps}$.
In the inset, we show the quality of the fit of the Debye spectral density~\cref{eq:debye:spectral:density}, using $N^I=19$ Lorentzians.
We checked that the small oscillations around the exact shape of $J^D(\omega)$ have no impact on the obtained dynamics.
Taking into account the additional $\delta$\hyp peak in $J^D(\omega)$, we performed computations at $T=77\mathrm{K}$.
The resulting, time\hyp dependent excitonic occupations are depicted in~\cref{fig:example:fmo:77k+peak}.
Note the additional $\delta$\hyp peak in the spectral density indicated by the grey dashed line of the inset.
The effect of the long\hyp term memory induced by the dissipationless additional mode immediately translates into the excitonic dynamics.
Compared to the Debye\hyp only case, we find long\hyp lived oscillations in the excitonic occupations, with an increased coherence time compared to Ref.~\onlinecite{nalbach2011exciton}, due to the exact treatment of the $\delta$\hyp peak.
\section{\label{sec:sf:benchmark}Benchmark}
\begin{figure}
    \centering
    \includegraphics[width=0.495\textwidth]{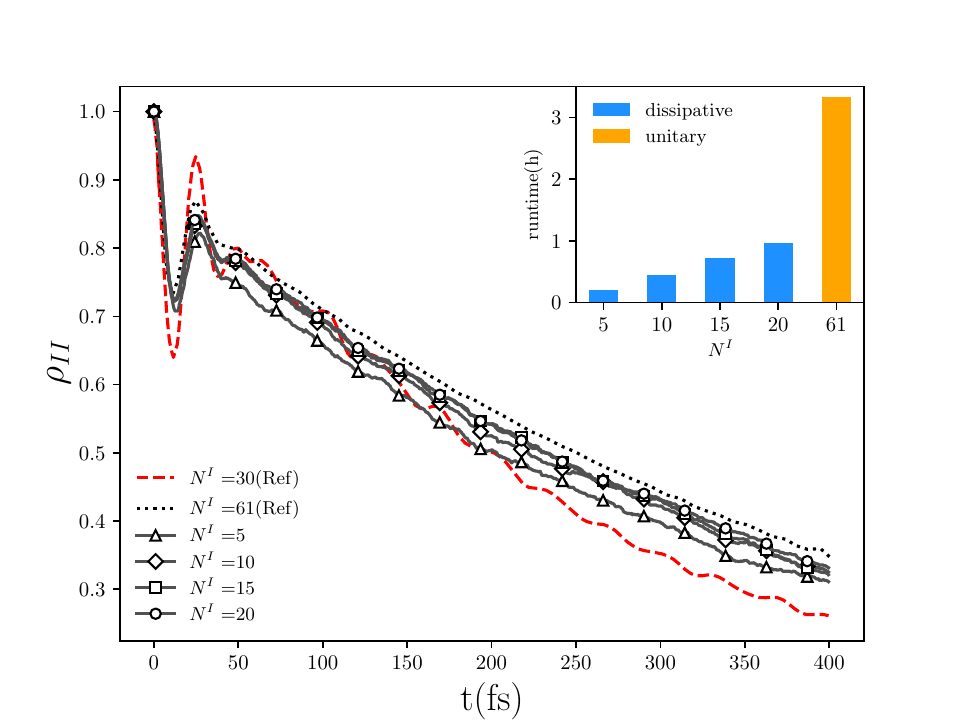}
    \caption{
        \label{fig:benchmark:sf}
 Convergence of the \gls{LE} population $\braket{\mathrm{LE}|\hat \rho|\mathrm{LE}}$ for the \gls{SF}\hyp example obtained from unitary evolution using $N^I=30$ (red dashed) and $N^I=61$(black dotted) modes, representing the most relevant vibrational modes~\cite{Xie2019-ce} compared to the dynamics obtained from~\gls{PS-HyB}\hyp\gls{MPS}.
 For the case of $N^I=30$ modes, the unitary dynamics exhibit stronger oscillations caused by finite\hyp size effects.
 Note that using~\gls{PS-HyB}\hyp\gls{MPS}, the unitary dynamics with $N^I=61$ modes is well\hyp reproduced but with a significantly smaller number of bosonic modes $N^I=10$.
 In the inset, we show a runtime comparison of the unitary dynamics (orange bar) with~\gls{PS-HyB}\hyp\gls{MPS} (blue bars).
 The computational speed up is nearly an order of magnitude, due to the drastically reduced number of bosonic modes required.
 }
\end{figure}
Besides ensuring the consistency of~\gls{PS-HyB}\hyp\gls{MPS}, we also performed benchmark computations to compare the efficiency of~\gls{PS-HyB}\hyp\gls{MPS} to that of standard, unitary dynamics, here at the example of~\gls{SF}(~\cref{sec:sf})\footnote{The case of the~\gls{FMO} complex is discussed in~\cref{app:dissipative:vs:unitary}.}.
In~\cref{fig:benchmark:sf}, we show the dynamics of the~\gls{LE} state (solid lines), computed using different numbers of $N^I=5, 10, 15, 20$ Lorentzians to approximate~\cref{eq:debye:spectral:density}.
The unitary dynamics was calculated taking into account $N^I=30$ (dashed line) and $N^I=61$ (dotted line) relevant vibrational modes where it has been shown that $N^I=61$ is required to faithfully describe the excitonic dynamics~\cite{Xie2019-ce}.
Due to the discrete nature of the vibrational environment in the unitary approach, finite\hyp size effects are observed that translate to stronger oscillations in the initial dynamics.
Instead,~\gls{PS-HyB}\hyp\gls{MPS} reproduces both, the initial and long\hyp time dynamics but with a much smaller number of bosonic modes.
This translates into a significant reduction of the runtime, which is shown in the inset\footnote{The calculations were performed on an Intel\textregistered Xeon\textregistered Gold 6130 CPU (2.10GHz) with 16 threads}.
Using only $N^I=10$ modes, the observed computational speedup is a factor of $\sim 7.5$.
We also note that the finite lifetime of the bosonic excitations generated by the dissipative evolution in~\gls{PS-HyB} continuously decreases both, the bond and effective bosonic Hilbert space dimensions, in the scope of the dynamics\footnote{See~\cref{app:dissipative:vs:unitary} for a comparison w.r.t. bond dimension and local Hilbert space dimension.}.
\section*{\label{sec:conclusion}Summary and Outlook}%
In this work, we introduced~\gls{PS-HyB}\hyp\gls{MPS}, a method that allows us to efficiently describe the influence of both continuous and discrete vibrational quantum baths on exciton dynamics.
To this end, we combine a Markovian\hyp embedding scheme with a pure\hyp state unraveling and state\hyp of\hyp the\hyp art tensor network techniques to represent and evolve bosonic baths with very large Hilbert space dimensions.
\Gls{PS-HyB}\hyp\gls{MPS} is capable of describing both short\hyp and long\hyp time memory effects in exciton dynamics at finite temperatures while requiring a much smaller amount of bosonic modes compared to commonly used unitary descriptions.
The resulting reduction in computational complexity translates into significantly lower numerical costs; for instance, we obtain a speedup of nearly an order of magnitude for a paradigmatic model of~\acrlong{SF}.
We believe that~\gls{PS-HyB}\hyp\gls{MPS} will be a valuable tool to tackle the description of intricated excitonic dynamics in vibrational environments, opening the possibility to describe mesoscopic environments efficiently by using wave function\hyp based time\hyp evolution schemes.
Moreover, the prospect of overcoming the current limitation of independent vibronic baths using non\hyp Markovian descriptions of the environment (see~\cref{app:markovian:embedding:vs:hops}) opens the possibility of increasing the computational efficiency of more complex problems, such as photo\hyp induced excitonic dynamics in the presence of coupled vibrational modes.
\section*{\label{sec:acknowledgement}Acknowledgments}%
The authors were grateful to Haibo Ma, Artur M. Lacerda and John Goold for fruitful discussions.
ZX, MM, US, and SP acknowledged support from the Munich Center for Quantum Science and Technology. %
US acknowledged funding by the Deutsche Forschungsgemeinschaft (DFG, German Research Foundation) under Germany’s Excellence Strategy-EXC-2111-390814868.
\bibliography{literature}%
\FloatBarrier
\onecolumngrid
\appendix 
\section{Numerical details}
\label{app:numerical:details}
\begin{table}
    \caption{\textbf{Parameters of 19 optimized modes for \gls{FMO}}} 
    \centering 
    \begin{tabular}{ccc}  
      \hline 
      $\omega/\mathrm{eV}$ & $\kappa/\mathrm{eV}$ & $\gamma/\mathrm{eV}$\\  
      \hline  
 2.30099e-03 & 6.88515e-04 & 9.50196e-04\\
 3.28506e-03 & 9.58415e-04 & 1.21617e-03\\
 4.33189e-03 & 1.20018e-03 & 1.47447e-03\\
 5.47123e-03 & 1.43038e-03 & 1.74525e-03\\
 6.72464e-03 & 1.65352e-03 & 2.04209e-03\\
 9.67419e-03 & 2.08386e-03 & 2.77137e-03\\
 8.11579e-03 & 1.87114e-03 & 2.37871e-03\\
 1.14380e-02 & 2.29190e-03 & 3.24048e-03\\
 1.34576e-02 & 2.49578e-03 & 3.81300e-03\\
 1.57994e-02 & 2.69633e-03 & 4.52527e-03\\
 1.85530e-02 & 2.89496e-03 & 5.42731e-03\\
 2.18399e-02 & 3.09376e-03 & 6.58909e-03\\
 2.58265e-02 & 3.29575e-03 & 8.10986e-03\\
 3.07445e-02 & 3.50484e-03 & 1.01299e-02\\
 3.69172e-02 & 3.72623e-03 & 1.28428e-02\\
 4.47954e-02 & 3.96608e-03 & 1.64867e-02\\
 5.49886e-02 & 4.23097e-03 & 2.12190e-02\\
 6.82436e-02 & 4.51903e-03 & 2.64008e-02\\
 8.52510e-02 & 4.74554e-03 & 2.68574e-02\\
      \hline 
      \label{table:7x19}
    \end{tabular}
  \end{table}
In this appendix, we provide additional details on the numerical methods used in the main text.
We represent the state of the excitonic and of the vibrational degrees of freedom as an \gls{MPS}~\cite{schollwock2011density}.
We use the \gls{PP} mapping ~\cite{10.21468/SciPostPhys.10.3.058} for the vibrational modes, which allows us to determine the optimal local dimensions for the vibrational modes at every time evolution step.
For all results presented in the main text, we allowed for a maximum local dimension $d_\mathrm{max}$ of $=16$.
For the time evolution, we used the \gls{TDVP} algorithm~\cite{haegeman2011time, haegeman2016unifying}.
Specifically, we employed the \gls{LSE-TDVP} \cite{PhysRevB.102.094315, grundner2023cooperpaired}, which is very efficient, especially for systems with large local dimensions, like the vibrational modes in our case.
This consists of single-site \gls{TDVP} steps with some maximal bond dimension $m$ alternated by expansions that increase the state's bond dimension by $m_e$.
Unless otherwise stated, we set $m=50$ and $m_e=10$.
For the \gls{TN} calculations, we used the \textsc{SyTen} toolkit \cite{hubig:_syten_toolk,hubig17:_symmet_protec_tensor_networ}.

We unraveled the Lindblad equation using the \gls{QJ} method (see also~\cref{app:dissipative:vs:unitary}) with a time step of $dt=5$ fs and either $N_\mathrm{trajs}=500$ or $N_\mathrm{trajs}=1000$ trajectories (see the figure captions).
The \gls{QJ} method was implemented in \textsc{evos}, a Python package for open quantum system dynamics that uses \textsc{SyTen} as the backend.
All computations were performed on an Intel Xeon Gold 6130 CPU (2.10GHz) with 16 threads.
For all singlet fission calculations, we fitted the spectral density using an equally spaced discretization.
Instead, for the \gls{FMO} dynamics we found it more convenient to use a non-equally-spaced discretization, as shown in~\cref{table:7x19}.
\section{Singlet fission with different number of trajectories}
\label{app:singlet:fission}
\begin{figure}
    \centering
    \includegraphics[width=0.7\textwidth]{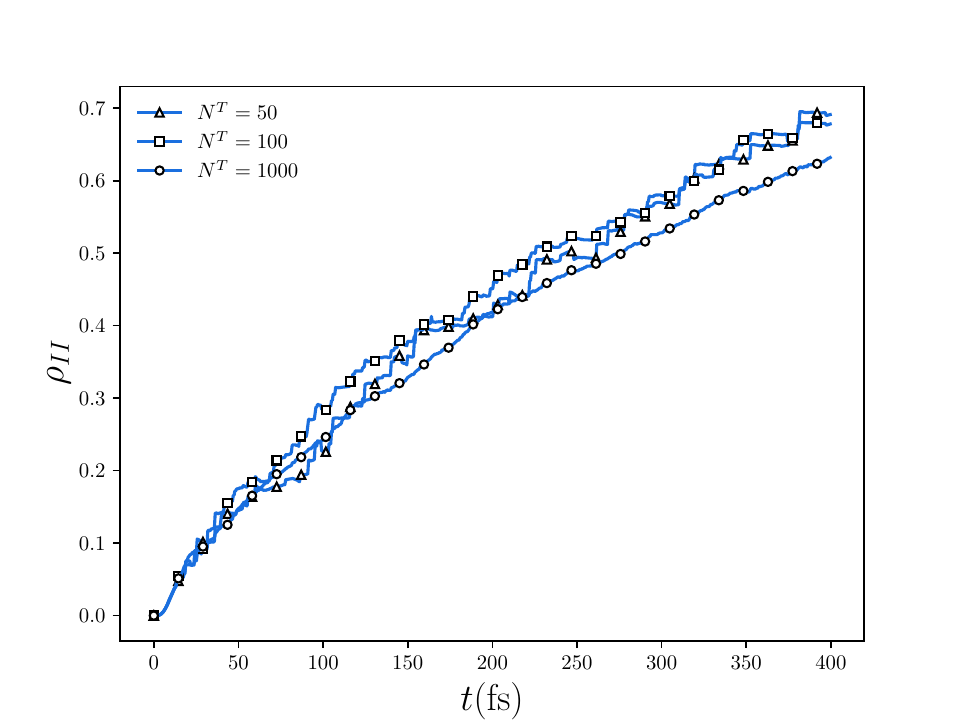}
    \caption{Population of \gls{TT} state in Markovian embedding dissipative dynamics(equally spaced $N^I = 10$) with different numbers of trajectories.}
    \label{fig:sf:trajs}
\end{figure}
In the main text, we computed the time evolution of a singlet fission model using Markovian embedding dissipative dynamics. 
We showed that accurate results for the dynamics can be obtained by replacing a large closed system ($N^I=61$) with a much smaller open system ($N^I=10$). 
The disadvantage of the dissipative approach is that one time evolution is replaced by an ensemble of trajectories.
However, we argue that this is only a minor disadvantage for the following reasons:
\begin{itemize}
    \item [(\romannumeral1)] Although reproducing smooth dynamics can require up to $\mathcal{O}(1000)$ trajectories, a very small number of trajectories already suffices to reproduce the qualitatively correct dynamics, i.e. the generation of \gls{TT} state( see \cref{fig:sf:trajs}).
 Therefore, a quick estimation of the dynamics can be easily achieved using this scheme instead of performing a fine discretization for each subregion of the spectrum \cite{Zheng2016-ju,xie2019time}. 
    \item [(\romannumeral2)] Each quantum trajectory is completely independent of the others.
 This enables a trivial, perfect parallelization over trajectories which enhances and simplifies high-performance computations.
    \item [(\romannumeral3)] For isolated many-body, strongly-correlated systems with bosonic \gls{DOFs}, \gls{TDVP} requires large bond dimensions $m$ and large local dimensions $d$. 
 Instead, as we demonstrate in~\cref{sec:un-vs-dis}, dissipatively coupling the vibrational modes to a bath suppresses both the bond dimension and the physical dimension.
\end{itemize}
\section{Comparing Markovian embedding and unitary evolution \label{sec:un-vs-dis}} 
\label{app:dissipative:vs:unitary}
\begin{figure}
    \centering
    \includegraphics[width=0.7\textwidth]{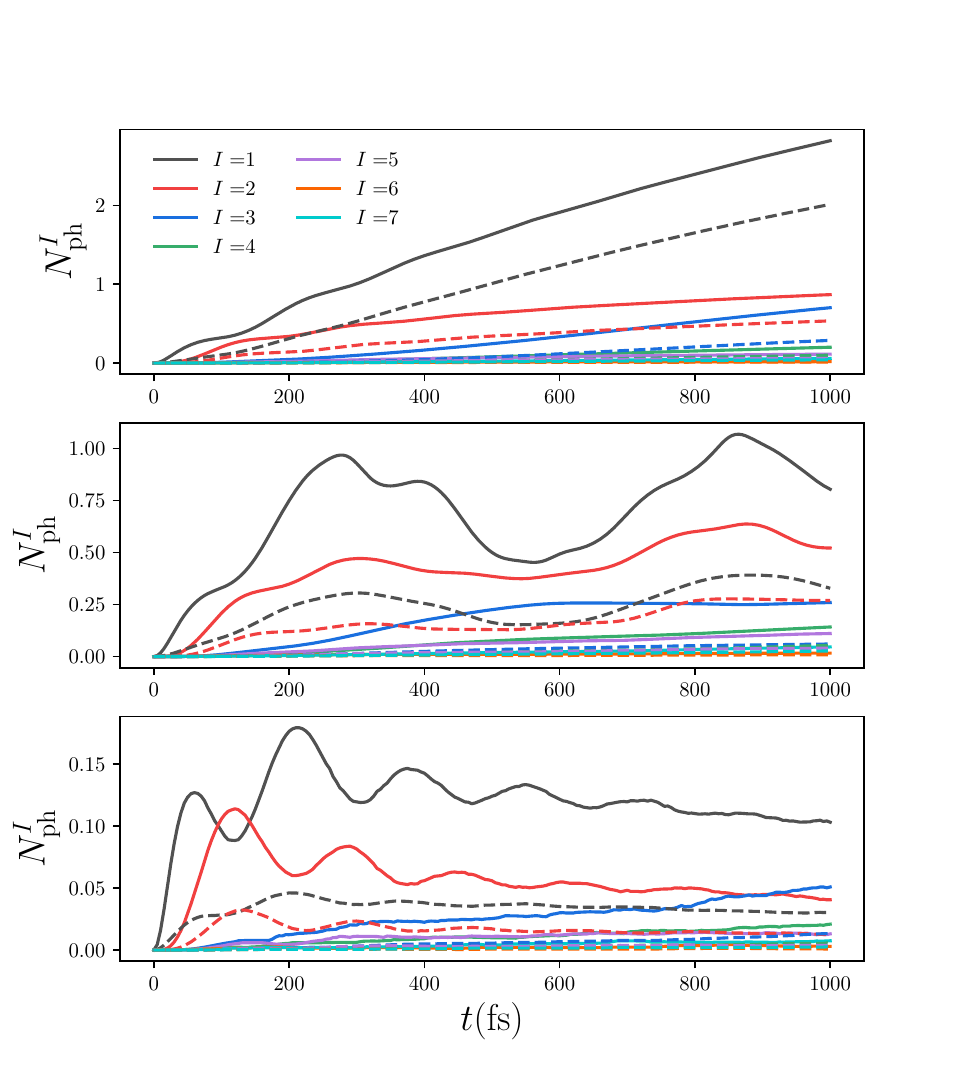}
    \caption{Total phononic occupation in the bath of each site($N_{ph}^I = \sum_k \langle \hat{b}^{I\dagger}_kb^I_k\rangle$) using equally spaced  $N^I = 30$ unitary evolution (upper panel), equally spaced $N^I = 6$ unitary evolution (middle panel) and equally spaced $N^I = 6$ Markovian embedding scheme (lower panel). 
 The phononic occupations of the unitary evolutions are larger than the ones of the Markovian embedding scheme by one order of magnitude.
 Here, solid lines indicate positive-frequency modes and dashed lines for negative-frequency modes in the \acrlong{TFD} scheme \cite{devega2015thermofield} (see~\cref{app:tfd:vs:unravel}).}
    \label{fig:phocc}
   \end{figure}
   
   \begin{figure}
    \centering
    \includegraphics[width=\textwidth]{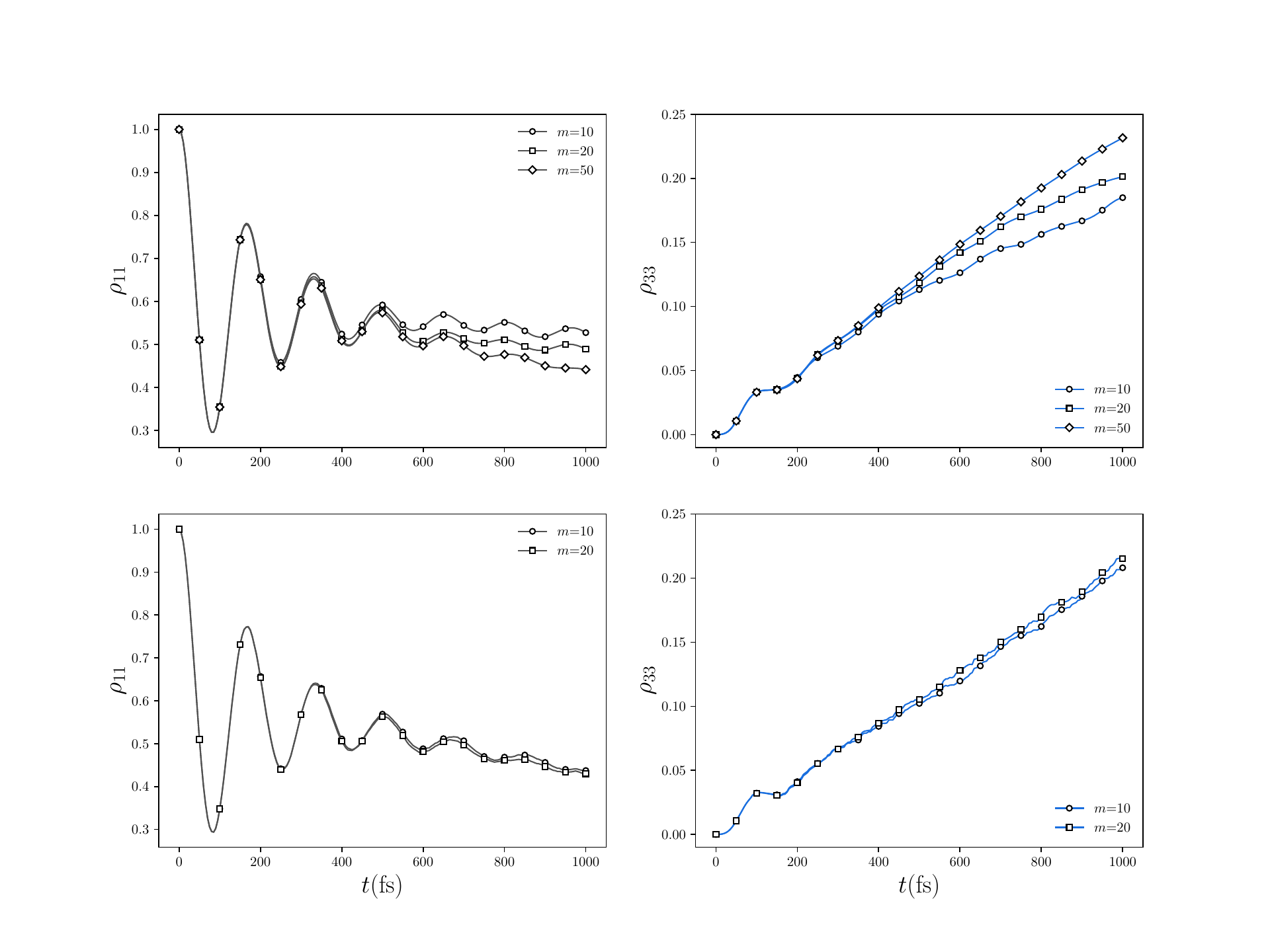}
    \caption{Bond dimension $m$ convergence of the unitary evolution (upper) and Markovian embedding (lower) using equally spaced  $N^I = 30$. 
 We show the populations of site $1$ (black, left panels) and site $3$ (blue, right panels). 
 The local dimensions of dissipative dynamics are fixed to $d = 2$.}
    \label{fig:bdim}
   \end{figure}
   
   \begin{figure}
    \centering
    \includegraphics[width=\textwidth]{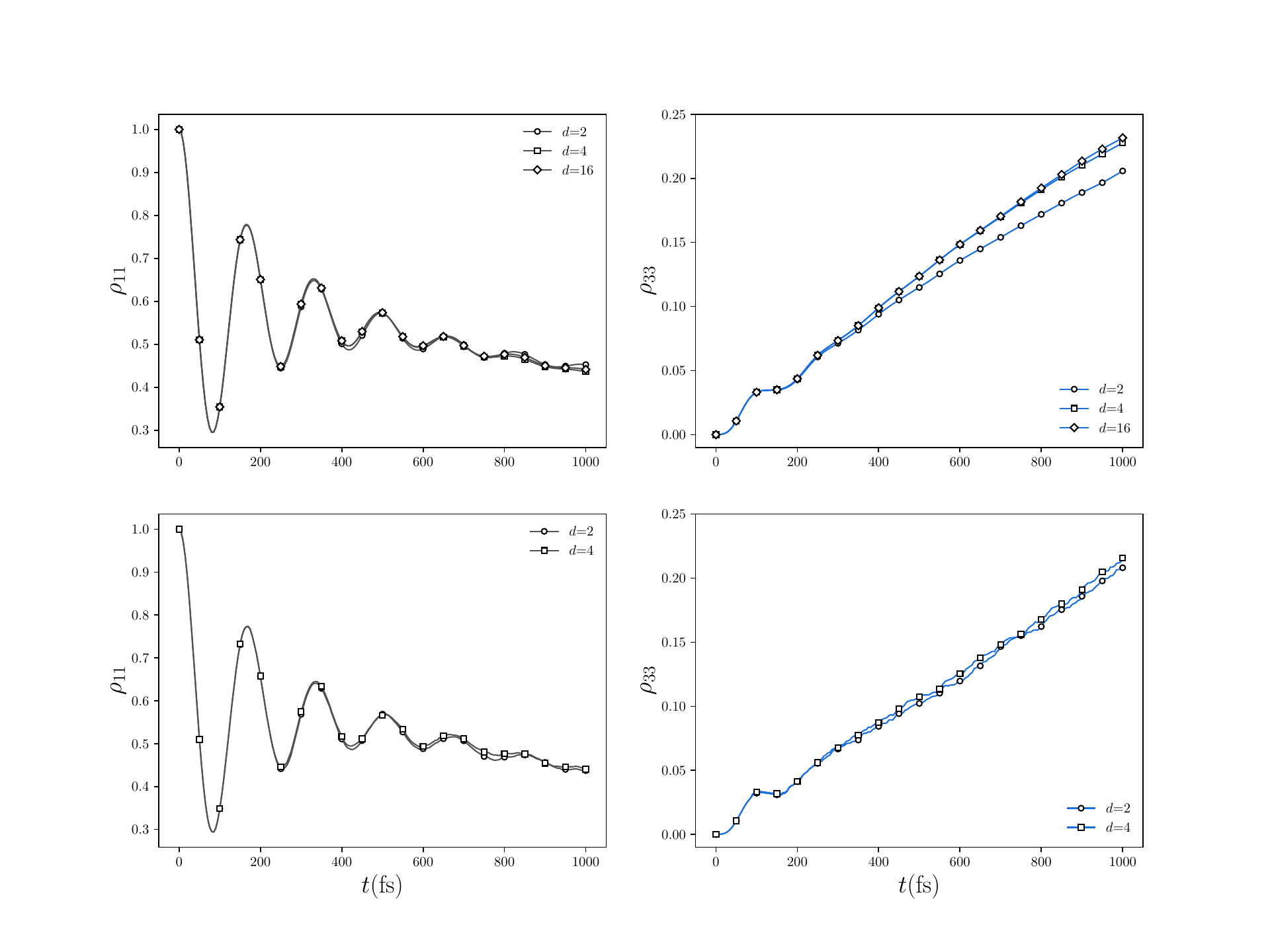}
    \caption{Local dimension $d$ convergence of the unitary evolution (upper) and Markovian embedding (lower) using equally spaced $N^I = 30$. 
 We show the populations of site $1$ (black, left panels) and site $3$ (blue, right panels). 
 The bond dimensions of dissipative dynamics are fixed to $m = 10$.}
    \label{fig:localdim}
   \end{figure}
In this section, we explain similarities and differences between unitary and (non-unitary)Markovian embedding schemes based on \gls{MPS} we performed in the main text.
First, they share the equivalent decomposition of a continuous spectral density into discrete modes.
For simplicity, we assume the spectral densities $J^I(\omega)$ are discretized into $N^I$ modes, equally spaced by $\Delta \omega$. 
The $k$\hyp{th} mode evolves with frequency $\omega_k$ and couples to the excitonic system via $g_k^I \hat{n}^I(\hat{b}^I_k + \hat{b}^{I\dagger}_k)/\sqrt{2}$. 
In the unitary scheme, the system and these discrete modes together form a large closed system where the coupling coefficients are
\begin{equation}
 g^I_k = \sqrt{\frac{2}{\pi}J^I(\omega_k)\Delta \omega} .
 \label{eq:discr:unitary}
\end{equation}
In the Markovian embedding scheme, the mesoscopic leads method \cite{imamog1994stochastic, Garraway1997, PhysRevB.86.125111, 10.1063/1.5000747, PhysRevA.101.050301, brenes2020tensor,lacerda2023quantum} decomposes the spectral densities into Lorentzians (with center $\tilde{\omega}_m$ and width $\gamma$). 
In analogy to~\cref{eq:discr:unitary}, the coupling between an excitonic mode $I$ and a vibrational mode with frequency $\tilde{\omega}_m$ is given by
\begin{equation}
 \tilde{g}^I_m = \sqrt{\frac{2}{\pi}J^I(\tilde{\omega}_m) \gamma}, 
\end{equation}
where the Lorentzian widths are chosen to be $\gamma = \tilde{\omega}_{m+1} - \tilde{\omega}_{m} = \Delta \tilde{\omega}$. 
The vibrational system is embedded in Markovian environments characterized by a dissipation strength $\gamma$.
Therefore, if the same discrete frequencies are chosen, these two schemes evolve the same excitonic-vibrational system either unitarily as a closed system where the rest of the continuous bath is completely ignored (namely, the broadening is completely absorbed into the discrete peak) or as an open system with the Markovian embedding. 

In the following, we compare the unitary dynamics to the dissipative Markovian-embedding dynamics unraveled with \gls{QJ}.
The evolution of a closed system with initial pure state $|\Psi(0)\rangle$ obeys the Schrödinger equation
\begin{equation}
 \frac{\partial}{\partial t}|\Psi(t)\rangle = -i \hat{H}|\Psi(t)\rangle,
 \label{eq:schroedinger}
\end{equation}
where $\hat{H}$ is the Hermitian Hamiltonian of the enlarged system discussed above (exctions and vibrational modes). 
The induced time-dependence is given by
\begin{equation}
 |\Psi(t+\delta t)\rangle = e^{-i\hat{H}\delta t}|\Psi(t)\rangle,
 \label{eq:unitary}
\end{equation}
and in practical numerical computations, one typically splits the unitary time evolution operator into several time steps
\begin{equation}
 |\Psi(t)\rangle = e^{-i\hat{H}\delta t} \cdots e^{-i\hat{H}\delta t}|\Psi(0)\rangle.
 \label{eq:unitary:split}
\end{equation}

In the Markovian embedding scheme, the evolution of the enlarged system obeys the Lindblad master equation
\begin{equation}
 \frac{\partial \hat{\rho}(t)}{\partial t} = -i \comm{\hat{H}_{\mathrm{eff}}}{\hat{\rho}(t)} + \sum_l \hat{L}^{\vphantom{\dagger}}_l \hat{\rho}(t)\hat{L}^\dagger_l,
 \label{eq:lindblad:qj:heff} 
\end{equation}
where $\hat{H}_{\mathrm{eff}} = \hat{H} -i/2 \sum_l \hat{L}^\dagger_l \hat{L}_l\equiv \hat{H} -i \hat{H}^{\prime}$. 
The \gls{QJ} method unravels the evolution of the density matrix $\hat{\rho}(t)$ into an ensemble of quantum trajectories \cite{daley2014quantum}.
Each trajectory follows a non\hyp{unitary} evolution, accompanied by quantum jumps happening at random times. 
At every timestep, with a probability of $1-\delta p$, a trajectory either evolves under a non-Hermitian Hamiltonian
\begin{equation}
 |\Psi(t+\delta t)\rangle = e^{-i\hat{H}_{\mathrm{eff}}\delta t}|\Psi(t)\rangle,
 \label{eq:qj:non-unitary}
\end{equation}
or undergoes a jump with a probability $\delta p$
\begin{equation}
 |\Psi(t+\delta t)\rangle = \hat{L}_l|\Psi(t)\rangle .
 \label{eq:qj:jump}
\end{equation}
Here, the probability of having a quantum jump is determined by the norm loss during non\hyp{unitary} evolution
\begin{equation}
 \delta p = 1 - \langle \Psi(t)|e^{+i\hat{H}^{\dagger}_{\mathrm{eff}}\delta t}e^{-i\hat{H}_{\mathrm{eff}}\delta t}|\Psi(t)\rangle .
 \label{eq:qj:loss}
\end{equation}
Up to first order in $\delta t$, the norm loss can be decomposed as
\begin{equation}
 \delta p \approx \delta t\langle \Psi(t)|i(\hat{H}_{\mathrm{eff}}-\hat{H}_{\mathrm{eff}}^{\dagger})|\Psi(t)\rangle = \delta t \sum_l \langle \Psi(t)|\hat{L}^{\dagger}_l \hat{L}_l|\Psi(t)\rangle \equiv \sum_l \delta p_{l}.
\end{equation}
Note that the Hermitian part of the effective Hamiltonian $\hat{H}$ is nothing but the Hamiltonian of the closed system in a unitary scheme. 
At zero temperature, the jump operators used in the mesoscopic leads method are 
\begin{equation}
 \hat{L}_l = \sqrt{\gamma}\;\hat{b}_k^I
 \label{eq:mesolead:tfd:jump}
\end{equation}

where $\gamma$ is the decay rate of the vibrational modes.
Thus, the probability of each decay channel is proportional to its phononic occupation
\begin{equation}
 \delta p_l = \gamma \delta t \langle\Psi(t)|\hat{b}^{I\dagger}_k\hat{b}^{I}_k|\Psi(t)\rangle.
\end{equation}
To compare with the unitary evolution more straightforwardly, we trotterize the evolution of $\hat{H}_{\mathrm{eff}}$ into the evolution of Hermitian $\hat{H}$ and anti-Hermitian part $-i\hat{H}^{\prime}$:
\begin{equation}
 |\Psi(t+\delta t)\rangle = e^{-i\hat{H}_{\mathrm{eff}}\delta t}|\Psi(t)\rangle \approx
 e^{-\hat{H}^{\prime}\delta t}e^{-i\hat{H}\delta t}|\Psi(t)\rangle .
 \label{eq:qj:trotter}
\end{equation}
Following \ref{eq:unitary:split}, the evolution is split into time steps
\begin{equation}
 |\Psi(t)\rangle = e^{-\hat{H}^{\prime}\delta t}e^{-i\hat{H}\delta t} \cdots e^{-\hat{H}^{\prime}\delta t}e^{-i\hat{H}\delta t}|\Psi(0)\rangle.
 \label{eq:qj:nonunitary:split}
\end{equation}
\ref{eq:qj:nonunitary:split} can be seen as an alternating sequence of imaginary evolution steps $e^{-\hat{H}^{\prime}\delta t}$ and real-time evolution steps $e^{-i\hat{H}\delta t}$. 
The imaginary {time} evolution can be expressed in terms of jump operators \ref{eq:mesolead:tfd:jump}
\begin{equation}
 e^{-\hat{H}^{\prime}\delta t} = e^{-\gamma\delta t/2\sum_k \hat{b}^{I\dagger}_k \hat{b}_k^I}.
 \label{eq:qj:imag}
\end{equation}
\ref{eq:qj:imag} indicates an exponential suppression of the phononic occupation with strength $\gamma\delta t$ at each unitary (real) time evolution step (see \cref{fig:phocc}). 
Such suppression results in a loss of norm \ref{eq:qj:loss} and can lower the local dimensions (\cref{fig:localdim}) and the bond dimensions (\cref{fig:bdim}) of \gls{TDVP} calculations.
Thus, an expensive unitary evolution can be substituted by much cheaper evolutions of multiple trajectories. 
Note that at finite temperatures, the creation operators are also the jump operators in the mesoscopic leads method. 
However, the prefactor of the annihilation jump operators is always larger than that of the creation operators, typically resulting in a decrease in the bond dimension and the local dimension compared to the unitary evolution at any temperature.
\section{Finite temperature: Thermofield doubling vs unraveling}
\label{app:tfd:vs:unravel}
For the simulations of the exciton dynamics of the \gls{FMO} complex discussed in the main text, we initialized the system in a tensor product state between a pure excitonic state and a thermal state of the vibrational system
\begin{equation}
 \hat{\rho}(0) = \ket{\psi(0)}_{\mathrm{ex}}\bra{\psi(0)}_{\mathrm{ex}} \otimes \hat{\rho}^\beta_\mathrm{vib}.
 \label{eq:app:initial:state}
\end{equation}
In order to represent the mixed state $\hat{\rho}^\beta_\mathrm{vib}$ as an \gls{MPS}, there are different possible strategies: for instance to double the Hilbert space for the vibrational degrees of freedom or to unravel the thermal state.

\begin{figure}
 \centering
 \includegraphics[width=0.7\textwidth]{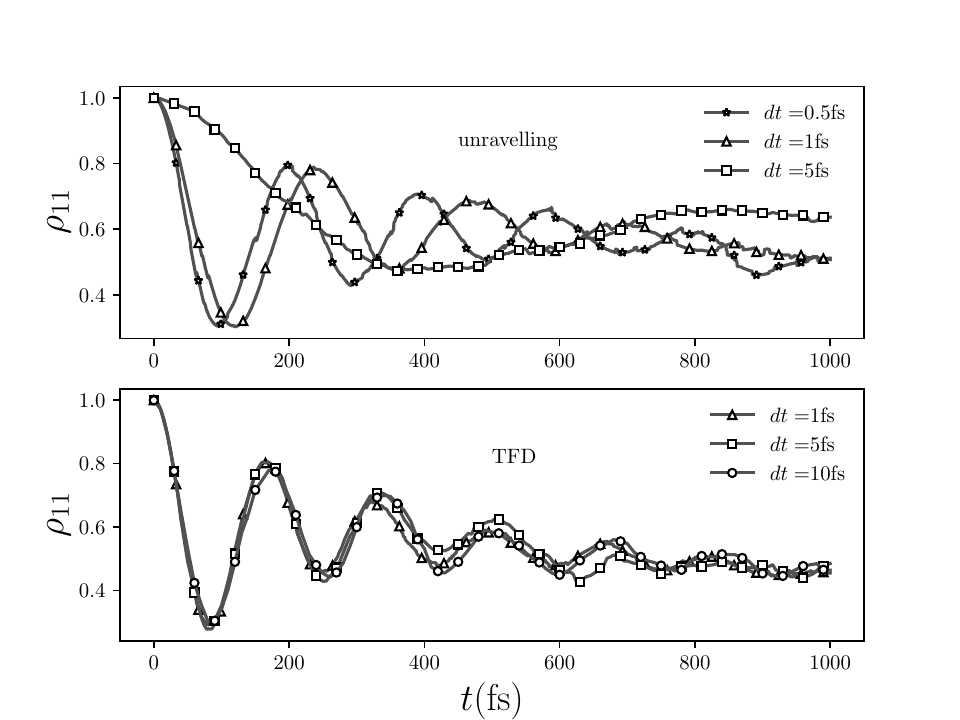}
 \caption{Thermal state unraveling vs. \acrfull{TFD}.
 The population of the first \gls{FMO} excitonic site at 77K was computed using Markovian embedding dissipative dynamics with 50 trajectories.
 The finite temperature is treated either by unraveling the thermal equilibrium state (upper panel) or via \gls{TFD} (lower panel).}
 \label{fig:unravel_vs_tfd}
\end{figure}

The first method, known as thermofield doubling, is widely used in \gls{MPS} calculations \cite{PhysRevB.72.220401, schollwock2011density}.
For Hamiltonians as the one considered in the main text, where the vibrational modes are described by independent harmonic oscillators and the coupling to the excitonic system is linear, de Vega and Bañuls in Ref. \onlinecite{devega2015thermofield} introduced a useful trick that simplifies the calculations. 
In short, the idea is to introduce a new set of bosonic operators $\hat{a}_k$, linked to the original operators $\hat{b}_k$ by a unitary transformation, that annihilates the thermal state $\hat{\rho}^\beta_\mathrm{vib}$.
By transforming the Hamiltonian to the new basis, the thermal state can be represented as a vacuum state.
This not only decreases the bond dimension of the initial state, but most importantly it reduces the number of quantum jumps from the environment to the vibrational modes, strongly improving the convergence of the ~\gls{QJ} method, as shown in~\cref{fig:unravel_vs_tfd}.
The second method consists in unraveling the thermal state $\hat{\rho}^\beta_\mathrm{vib} = \sum_k p(k) \ket{\psi^k}_\mathrm{vib} \bra{\psi^k}_\mathrm{vib} $, by sampling pure states $\ket{\psi^k}_\mathrm{vib}$ according to the probability distribution $p(k)$. 
This additional sampling of the initial state, which adds to the sampling of the \gls{QJ} method, increases the total number of trajectories required to converge the simulations.
Since the vibrational degrees of freedom are described by independent harmonic oscillators, the probability $p^I_m(n)$ of finding $n$ vibrational excitations for the $m$\hyp th mode relative to the $I$\hyp th exciton is 
\begin{equation}
    p^I_m(n) = \frac{e^{-\beta \omega ^I_m n}}{\Tr(e^{-\beta \omega^I_m n})}\;.
    \label{eq:fmo:sampling}
\end{equation}
%
In~\cref{fig:unravel_vs_tfd} we compare the performance of the thermofield doubling against the unraveling of the thermal state.
We simulate the occupation of the first excitonic site of the \gls{FMO} at $77K$ using 50 trajectories for different timesteps $dt$.
It can be seen that while the thermofield method yields good results for $dt=5 $fs the unraveling method is not converged even for a ten times smaller timestep. 
This comes from the fact that at $77K$ the probability of quantum jumps happening between the thermal environment and the vibrational modes is much higher than at zero temperature, and the increased number of jumps strongly slows down the real\hyp{time} evolution. 
Therefore, the mapping of the thermal state to a vacuum state~\cite{devega2015thermofield} is crucial.

\section{Markovian embedding vs non-Markovian methods}
\label{app:markovian:embedding:vs:hops}
\begin{figure}
    \centering
    \includegraphics[width=0.7\textwidth]{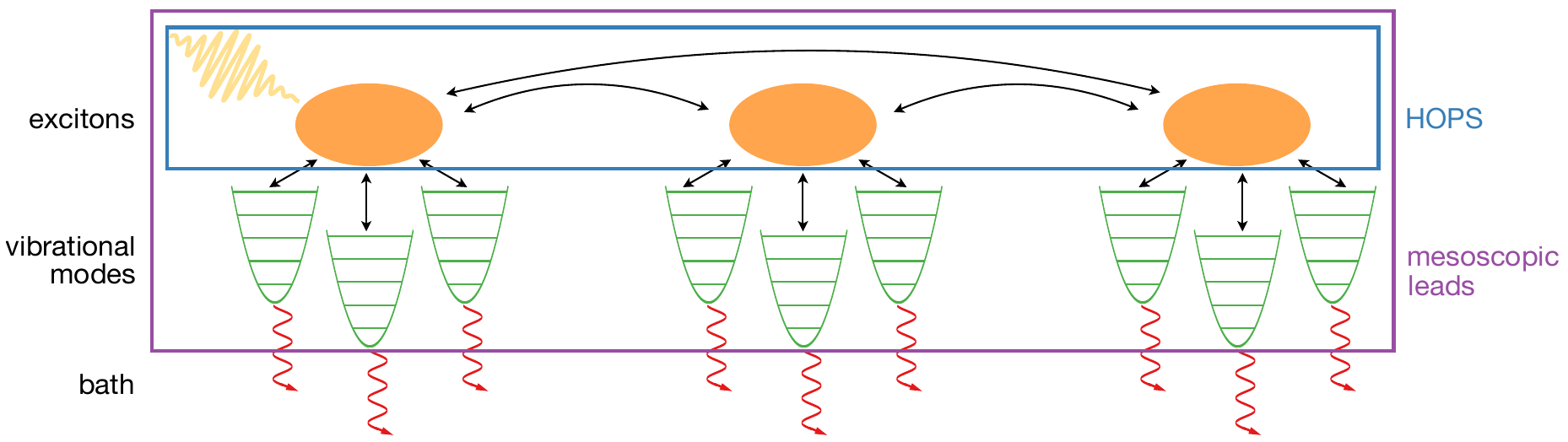}
    \caption{Two equivalent descriptions of exciton dynamics in a vibrational environment.
 Non-Markovian methods like \gls{HOPS} consider only the exciton subsystem explicitly, while Markovian-embedding techniques as the mesoscopic leads method include also the vibrational modes.}
    \label{fig:hops:mesoleads}
\end{figure}
In the main text, we have described the photo-induced dynamics of exciton systems strongly coupled to vibrational degrees of freedom by applying the Markovian-embedding-based mesoscopic leads method~\cite{imamog1994stochastic, Garraway1997, PhysRevB.86.125111, 10.1063/1.5000747, PhysRevA.101.050301, brenes2020tensor,lacerda2023quantum}. 

This consists of evolving the full excitonic-vibrational Hamiltonian exactly and damping the vibrational modes through a Lindblad dissipator.
Here we outline an alternative, non-Markovian description and explore its connection to the mesoscopic leads method.

Instead of evolving the combined excitonic-vibronic state $\ket{\Psi}$ with the full Hamiltonian 
\begin{equation}
    \hat{H} = \hat{H}_{\mathrm{ex}} + \hat{H}_{\mathrm{vib}} + \hat{H}_{\mathrm{ex-vib}},
    \label{eq:app:total:ham}
\end{equation}
non-Markovian techniques allow us to deal directly only with $\hat{H}_{\mathrm{ex}}$ and consider the impact of the vibrational modes implicitly.
The two possible system-environment bipartitions are shown in~\cref{fig:hops:mesoleads}
For the sake of simplicity, let us consider the interaction Hamiltonian studied in the main text (although more general interactions are possible~\cite{suess2014hierarchy}):
\begin{equation}
    \hat{H}_{\mathrm{ex-vib}} = \sum_{m,I} g^I_m \hat{n}^I ( \hat{b}^I_m + \hat{b}^{I \dagger}_m )/\sqrt{2},
    \label{eq:app:ham:ex:vib}
\end{equation}
where $g$ is the coupling strength, $\hat{n}$ is the excitonic number operator, and $\hat{b}^\dagger$ ($\hat{b}$) creates (annihilates) one vibrational mode. 
Differently from the mesoscopic leads approach, one describes the system-bath interaction by working in the time domain and considering the bath correlation function
\begin{equation}
    \alpha^I(t-t') = 
   \int_0^\infty \mathrm{d}\omega J^I(\omega) \left [ \coth(2\beta \omega)\cos(\omega(t-t')) -i \sin(\omega(t-t'))  \right ].
    \label{eq:app:bath:corr:funct}
\end{equation}
Here $J^I(\omega)$ is the frequency-dependent bath spectral density, the superscript $I$ labels the different baths, and $\beta$ is the inverse bath temperature.
Instead of of fitting $J^I(\omega)$ with Lorentzians, we  approximate $\alpha^I(t-t')$ with a sum of complex exponentials, for instance via the Laplace-Pade method~\cite{molecules26164838}, as:
\begin{equation}
      \alpha^I(t-t') \approx \sum_m ^N g^I_m e^{\gamma^I_m \abs{t-t'} -i \omega^I_m (t-t') },
\end{equation}
where $\gamma$ and $\omega$ are the damping rate and the vibrational frequency, respectively.

In the following, we will denote a pure state on the combined exciton-phonon system by $\ket{\Psi}$ and for the excitons only with $\ket{\psi}$.
In analogy to the Markovian case~\cite{gardiner00}, it is possible to formulate a stochastic equation of motion for pure-state trajectories $q$ (see main text), so that the ensemble average 
\begin{equation}
    \mathcal{E} [\ket{\psi^q(t)} \bra{\psi^q(t)}] = \hat{\rho}^\mathrm{ex}(t)
\end{equation}
yields the correct time-evolved mixed state $\hat{\rho}^{\mathrm{ex}}$ for the excitons.

Here, $z(t)$ is a time-dependent complex colored noise with mean zero and correlations $\mathcal{E} [z(t) z^*(t')] = \alpha(t-t')$ and $\mathcal{E} [z(t) z(t')] = 0$~\cite{PhysRevA.71.023812}.
The time evolution for each trajectory is described by the so-called \gls{NMQSD}
\begin{equation}
    \partial_t \ket{\psi(t)} = -i \hat{H}_\mathrm{ex} \ket{\psi(t)} + \sum_{m,I} g^I_m z^{I*}_m(t) \hat{n}^I \ket{\psi(t)} - \sum_{m,I} g^I_m \hat{n}^I \int_0^t \mathrm{d}t' \alpha^{I*}_m(t-t') \frac{\delta \ket{\psi(t)}}{\delta z^{I*}_m(t')},
    \label{eq:app:nmqsd}
\end{equation}
where $\delta/\delta z^{I*}_m$ indicates a functional derivative.
In practice, solving this equation directly is unfeasible because the third term is time non-local.

An efficient strategy for solving~\cref{eq:app:nmqsd} was put forward by Suess, Eisfeld, and Strunz, who introduced the \gls{HOPS} method~\cite{suess2014hierarchy}.
The main idea of \gls{HOPS} consists of approximating the time non-local ~\cref{eq:app:nmqsd} with a hierarchy of time-local equations, which are simpler to solve.
In~\cite{PhysRevLett.128.063601,Gao2022}, the authors showed that these hierarchies can be mapped to bosonic modes, facilitating their representation as \glspl{MPS}.
For every trajectory $q$, the time evolution for the state on the combined excitonic-auxiliary moded $\ket{\Psi}$ reduces to 
\begin{equation}
    \partial_t \ket{\Psi(t)} = \hat{H}^q_{\mathrm{eff}}(t) \ket{\Psi},
\end{equation}
where $\hat{H}^q_{\mathrm{eff}}(t)$ is a non-Hermitian, time-dependent effective Hamiltonian~\cite{PhysRevLett.128.063601, Hartmann2017-ke}.

In~\cite{moroder2023stable}, a systematic comparison of a Markovian-embedding \gls{QJ} description with \gls{HOPS} was carried out for the Hubbard-Holstein model, which describes the interaction of electrons with locally-coupled Einstein phonons.
Interestingly, \gls{HOPS} was found to perform best in the presence of strong dissipation (i.e. strong damping $\gamma$), where the non-Markovian effects and thus the hierarchy depth are strongly suppressed.
Instead, \gls{QJ} was the method of choice in the case of weak dissipation, which is because for $\gamma \to 0$ it reduces to simple unitary dynamics for the enlarged system.
Based on this, we believe that in the context of optically-induced exciton dynamics, \gls{HOPS} can prove to be beneficial compared to \gls{QJ} when large damping rates $\gamma$ are considered, i.e. either when few vibrational modes are used to discretize the bath or for large bath temperatures.
A comparison between the mesoscopic leads and the \gls{HOPS} methods for these setups, as well as their extension to correlated baths, will be the topic of future investigations.
\end{document}

%% file: header.tex
\usepackage{lineno}
\usepackage{graphicx}  
\usepackage{dcolumn}   
\usepackage{bm}        
\usepackage{amssymb}   
\usepackage{amsmath}
\usepackage{blkarray, multirow, graphicx, diagbox, color, xcolor, colortbl}
\usepackage[caption=false]{subfig}
\usepackage{bbm, bbold}
\usepackage{ifthen}
\usepackage[colorlinks, linkcolor = blue, citecolor = blue, filecolor = black, urlcolor = blue]{hyperref}
\usepackage{xkeyval}
\usepackage{moreverb}
\usepackage{rotating}
\usepackage{slashbox}
\usepackage{xspace}
\usepackage{nicefrac}
\usepackage[]{units}
\usepackage{physics}
\usepackage{braket}
\usepackage[inline]{enumitem}
\usepackage{tabto}
\usepackage{listings}
\usepackage{xstring}
\usepackage{glossaries}
\usepackage{etoolbox}
\usepackage[makeroom]{cancel}
\usepackage{hyphenat}
\usepackage{booktabs}
\usepackage{placeins}

\glsdisablehyper

\def\ReplaceStr#1{%
	\IfSubStr{#1}{p}{%
		\StrSubstitute{#1}{p}{.}}{#1}}

\newcommand\subfigref[1]{\protect\subref{#1}}

\usepackage[customcolors]{hf-tikz} 
\usepackage{tikz}
\usepackage{calc}
\usetikzlibrary{external}
\graphicspath{{figures/autogen/}}
\usepackage{pgffor}
\usepackage{pgfplots}
\pgfplotsset{compat=newest}
\usepackage{pgfplotstable}
\usepgfplotslibrary{groupplots}
\usepgfplotslibrary{fillbetween}

\tikzstyle{n} = [draw,shape=ellipse,minimum size=1.5em,inner sep=0pt,fill=white!20, minimum width=2.5em]
\tikzstyle{Init} = [n,color=green,fill=green!20,text=black]
\tikzstyle{Fin} = [n,color=red,fill=red!20,text=black]
\tikzstyle{Ghost} = [minimum size=1.5em,inner sep=0pt,color=white,text=black]
\tikzstyle{Multiple} = [draw,shape=rect,minimum size=2em,inner sep=0pt]

\tikzstyle{ghostA} = [text=red!70,thick, minimum size=2*(5pt-\pgflinewidth), inner sep=0pt, outer sep=0pt]
\tikzstyle{ghostB} = [text=blue!70,thick, minimum size=2*(5pt-\pgflinewidth), inner sep=0pt, outer sep=0pt]
\tikzstyle{siteA} = [draw=red!70,circle,thick, minimum size=2*(5pt-\pgflinewidth), inner sep=0pt, outer sep=0pt]
\tikzstyle{siteB} = [draw=blue!70,circle,thick, minimum size=2*(5pt-\pgflinewidth), inner sep=0pt, outer sep=0pt]
\tikzstyle{operatorA} = [cross out, draw=red!70, thick, minimum size=2*(5pt-\pgflinewidth), inner sep=0pt, outer sep=0pt]
\tikzstyle{operatorB} = [cross out, draw=blue!70, thick, minimum size=2*(5pt-\pgflinewidth), inner sep=0pt, outer sep=0pt]

\tikzstyle{site} = [circle,thick,inner sep=0.2pt,minimum width=1.25em,font=\footnotesize,draw=blue!50!white,fill=blue!15!white,text opacity=1]
\tikzstyle{unsite} = [circle, outer sep=0pt,inner sep=0.2pt,minimum width=1.25em]
\tikzstyle{ghost} = []
\tikzstyle{op} = [regular polygon, regular polygon sides=4, draw=orange!50, fill=orange!20, thick, inner sep=0.2pt, minimum width=1.25em, minimum height=1.5em,font=\footnotesize]
\tikzstyle{ld} = [inner sep=1pt, font=\small]
\tikzstyle{intersite} = [regular polygon, regular polygon sides=4, shape border rotate= 45, draw=black!50,fill=black!20,thick,inner sep=0pt,minimum width=1.5em]
\usetikzlibrary{decorations.pathreplacing, calligraphy}

\definecolor{colorA}{rgb} {0.48,0,0.5275}
\definecolor{colorB}{rgb} {0.11,0.663,0.51}
\definecolor{colorC}{rgb} {0.3373,0.7059,0.9137}
\definecolor{colorD}{rgb} {0.902,0.8735,0.1}
\definecolor{colorE}{rgb} {0.9451,0.902,0.3255}
\definecolor{colorF}{rgb} {0.3373,0.3255,0.902}
\definecolor{colorG}{rgb} {0.9451,0.3255,0.3373}
\definecolor{colorH}{rgb} {0.11,0.3255,0.3373}

\usetikzlibrary
{
	calc,
	decorations,
	pgfplots.patchplots,
	plotmarks,
	patterns,
	positioning,
	petri,
	arrows,
	intersections,
	decorations.markings,
	backgrounds,
	fit,
	matrix,
	graphs,
	shapes.geometric,
	decorations.pathreplacing, 
	decorations.pathmorphing,
	shapes.misc,
	shapes.multipart,
	shapes,
	through,
	tikzmark
}

\pgfplotsset{
	cycle from colormap manual style/.style={
		x=3cm,y=10pt,ytick=\empty,
		stack plots=y,
		every axis plot/.style={line width=2pt},
	},
}

\tikzset{->-/.style={decoration={
			markings,
			mark=at position .5 with {\arrow{>}}},postaction={decorate}}}

\tikzset{-<-/.style={decoration={
			markings,
			mark=at position .5 with {\arrow{>}}},postaction={decorate}}}

\tikzstyle{orientedsnake} = [
decorate, 
decoration={snake},
->
]  
\tikzstyle{orientedshortarrow} = [
decoration={markings,
	mark=at position .33 with {\arrow{>}}},
postaction={decorate}
]  
\tikzstyle{orientedlongarrow} = [
decoration={markings,
	mark=at position .67 with {\arrow{>}}},
postaction={decorate}
]
\tikzset{dbl/.style={double,
		double equal sign distance,
		-implies,
		shorten >=10pt,
		shorten <=10pt}}
\tikzset{
	between/.style args={#1 and #2}{
		at = ($(#1)!0.5!(#2)$)
	}
}
\pgfmathdeclarefunction{linearFct}{2}%
{%
	\pgfmathparse{#1*x+#2}%
}
\pgfmathdeclarefunction{logFct}{2}%
{%
	\pgfmathparse{#1*log10(x)+#2}%
}
\pgfmathdeclarefunction{algebraicFct}{4}%
{%
	\pgfmathparse{#1*x^(#2)+#3/x+#4}%
}
\pgfmathdeclarefunction{specialFct}{4}%
{%
	\pgfmathparse{#1*x^(#2)*exp(#3/x)+#4}%
}
\pgfmathdeclarefunction{pointmetalog}{3}%
{%
	#1
}%

\newcommand{\nodagger}[0]{{\phantom{\dagger}}}
\newcommand{\noprime}[0]{{\phantom{\prime}}}

\newcommand{\discussive}[1]{\textcolor{black}{#1}}

\pgfmathdeclarefunction{peierlspotential}{6}%
{
	\pgfmathparse
	{
		#2 / (#3 * #5) 
		* sin(deg(#1 * #3 - #3 * #5 * (x)))
		* exp(-( ( #1 - #5 * ((x) - #4) ) )^2.0 / ( #6^2.0 ) )
	}%
}

\pgfmathdeclarefunction{theoreticalLimit}{2}
{
	\pgfmathparse
	{
		#2*(#1-1.0)*(#1*1.0-(#2*(#1-1.0)+1.0))/(#1*1.0)+0*x
	}
}

\newboolean{buildtikzpics}
\setboolean{buildtikzpics}{false}

\usepackage{ulem}
\usepackage{cleveref}
\Crefname{appendix}{Appendix}{Appendices}
\Crefname{equation}{Equation}{Equations}
\Crefname{figure}{Figure}{Figures}
\Crefname{section}{Section}{Sections}
\Crefname{tabular}{Tabular}{Tabulars}
\crefname{appendix}{App.}{Apps.}
\crefname{equation}{Eq.}{Eqs.}
\crefname{figure}{Fig.}{Figs.}
\crefname{section}{Sec.}{Secs.}
\crefname{tabular}{Tab.}{Tabs.}

\tikzset{>=stealth}

\pgfplotsset{
	compat=1.12,
	/pgf/declare function={
		cos2(\x) = cos(deg(\x*pi));
		g_plus(\x,\t,\s) = (\s-sqrt(\s^2+\t^2*cos2(\x)^2))/(\t*cos2(\x));
		g_minus(\x,\t,\s) = (\s+sqrt(\s^2+\t^2*cos2(\x)^2))/(\t*cos2(\x));
		gp(\x,\t) = ((\x - sqrt(\x^2 + \t^2))/\t); 
		gm(\x,\t) = ((\x + sqrt(\x^2 + \t^2))/\t);
	} 
}

\pgfplotsset{%
	colormap={parula}{%
		rgb=(0.2081,0.1663,0.5292)rgb=(0.2116,0.1898,0.5777)rgb=(0.2123,0.2138,0.627)
		rgb=(0.2081,0.2386,0.6771)rgb=(0.1959,0.2645,0.7279)rgb=(0.1707,0.2919,0.7792)
		rgb=(0.1253,0.3242,0.8303)rgb=(0.0591,0.3598,0.8683)rgb=(0.0117,0.3875,0.882)
		rgb=(0.006,0.4086,0.8828) rgb=(0.0165,0.4266,0.8786)rgb=(0.0329,0.443,0.872)
		rgb=(0.0498,0.4586,0.8641)rgb=(0.0629,0.4737,0.8554)rgb=(0.0723,0.4887,0.8467)
		rgb=(0.0779,0.504,0.8384) rgb=(0.0793,0.52,0.8312)  rgb=(0.0749,0.5375,0.8263)
		rgb=(0.0641,0.557,0.824)  rgb=(0.0488,0.5772,0.8228)rgb=(0.0343,0.5966,0.8199)
		rgb=(0.0265,0.6137,0.8135)rgb=(0.0239,0.6287,0.8038)rgb=(0.0231,0.6418,0.7913)
		rgb=(0.0228,0.6535,0.7768)rgb=(0.0267,0.6642,0.7607)rgb=(0.0384,0.6743,0.7436)
		rgb=(0.059,0.6838,0.7254) rgb=(0.0843,0.6928,0.7062)rgb=(0.1133,0.7015,0.6859)
		rgb=(0.1453,0.7098,0.6646)rgb=(0.1801,0.7177,0.6424)rgb=(0.2178,0.725,0.6193)
		rgb=(0.2586,0.7317,0.5954)rgb=(0.3022,0.7376,0.5712)rgb=(0.3482,0.7424,0.5473)
		rgb=(0.3953,0.7459,0.5244)rgb=(0.442,0.7481,0.5033) rgb=(0.4871,0.7491,0.484)
		rgb=(0.53,0.7491,0.4661)  rgb=(0.5709,0.7485,0.4494)rgb=(0.6099,0.7473,0.4337)
		rgb=(0.6473,0.7456,0.4188)rgb=(0.6834,0.7435,0.4044)rgb=(0.7184,0.7411,0.3905)
		rgb=(0.7525,0.7384,0.3768)rgb=(0.7858,0.7356,0.3633)rgb=(0.8185,0.7327,0.3498)
		rgb=(0.8507,0.7299,0.336) rgb=(0.8824,0.7274,0.3217)rgb=(0.9139,0.7258,0.3063)
		rgb=(0.945,0.7261,0.2886) rgb=(0.9739,0.7314,0.2666)rgb=(0.9938,0.7455,0.2403)
		rgb=(0.999,0.7653,0.2164) rgb=(0.9955,0.7861,0.1967)rgb=(0.988,0.8066,0.1794)
		rgb=(0.9789,0.8271,0.1633)rgb=(0.9697,0.8481,0.1475)rgb=(0.9626,0.8705,0.1309)
		rgb=(0.9589,0.8949,0.1132)rgb=(0.9598,0.9218,0.0948)rgb=(0.9661,0.9514,0.0755)
		rgb=(0.9763,0.9831,0.0538)
	}
}

\newcommand{\splittedTableHeader}[2]%
{%
	\tikzset{external/export next=false}%
	\begin{tikzpicture}%
		\node[anchor=south west, inner sep = 0, outer sep = 0] (n) at (0,0) {\tiny #1};%
		\node[anchor=north east, inner sep = 0, outer sep = 0] (d) at (0,0) {\tiny #2};%
		\node[fit = (n) (d), inner sep = 0, outer sep = 0] (frame) {};%
		\draw[-] (frame.north west) -- (frame.south east);%
	\end{tikzpicture}%
}%

\newcommand{\centeredSplittedTableHeader}[2]%
{%
	\noindent\parbox[c]{\widthof{\splittedTableHeader{#1}{#2}}}%
	{\splittedTableHeader{#1}{#2}}%
}%

%% file: acronyms.tex
\newacronym[shortplural={MPS}]{MPS}{MPS}{matrix\hyp product state}
\newacronym{TN}{TN}{tensor network}
\newacronym{DMRG}{DMRG}{density matrix renormalization group}
\newacronym{MCTDH}{MCTDH}{multi\hyp configurational time\hyp dependent Hartree}
\newacronym{ML-MCTDH}{ML-MCTDH}{multi\hyp layer multi\hyp configurational time\hyp dependent Hartree}
\newacronym{TDVP}{TDVP}{time\hyp dependent variational principle}
\newacronym{LSE-TDVP}{LSE-TDVP}{local subspace expansion time\hyp dependent variational principle}
\newacronym{GSE-TDVP}{GSE-TDVP}{global subspace expansion time\hyp dependent variational principle}
\newacronym{QSD}{QSD}{quantum state diffusion}
\newacronym{QJ}{QJ}{quantum jumps}
\newacronym{TFD}{TFD}{thermofield doubling}
\newacronym{PP}{PP}{projected-purification}
\newacronym[shortplural={PP-MPS}]{PP-MPS}{PP-MPS}{projected\hyp purified matrix\hyp product states}
\newacronym{NMQSD}{NMQSD}{non\hyp Markovian quantum state diffusion}
\newacronym{HOPS}{HOPS}{hierarchy of pure states}
\newacronym{DOFs}{DOFs}{degrees of freedom}
\newacronym{HEOM}{HEOM}{hierarchical equation of motion}
\newacronym{ADT}{ADT}{augmented density tensor}
\newacronym{QUAPI}{QUAPI}{quasi\hyp adiabatic path integral}
\newacronym{FMO}{FMO}{Fenna\hyp Matthews\hyp Olson}
\newacronym{SF}{SF}{singlet fission}
\newacronym{LE}{LE}{locally excited}
\newacronym{CT}{CT}{charge transfer}
\newacronym{TT}{TT}{two\hyp triplet}
\newacronym{PS-HyB}{PS-HyB}{pure\hyp state unraveled hybrid\hyp bath}
\newacronym{1RDM}{1RDM}{single\hyp site reduced density\hyp matrix}
\newacronym{DAMPF}{DAMPF}{dissipation\hyp assisted matrix\hyp product factorization}